\documentclass[prx,aps,superscriptaddress,footinbib,twocolumn]{revtex4-2}

\usepackage{mathrsfs}
\usepackage{amsfonts}
\usepackage{amssymb}
\usepackage{amsmath}
\usepackage{graphicx}
\usepackage[usenames,dvipsnames]{color}
\usepackage[colorlinks=true,citecolor=blue,linkcolor=magenta]{hyperref}
\usepackage{ulem}
\usepackage{lmodern}
\usepackage{amsthm}
\usepackage{dsfont}
\usepackage{natbib}
\usepackage{cancel}

\usepackage[noend]{algpseudocode}
\usepackage{algorithm}

\newtheorem{theorem}{Theorem}

\newcommand{\figpath}{.}

\newcommand{\Tr}{\mathrm{Tr}}
\newcommand{\norm}[1]{\Vert #1 \Vert}

\newcommand{\abs}[1]{\vert #1 \vert}

\newcommand{\ket}[1]{\vert{ #1 }\rangle}
\newcommand{\bra}[1]{\langle{ #1 }\vert}
\newcommand{\ketbra}[2]{\vert #1 \rangle \langle #2 \vert}

\newcommand{\mean}[1]{\langle #1 \rangle}

\newcommand{\bfC}{\boldsymbol{C}}
\newcommand{\bfT}{\boldsymbol{T}}

\newcommand{\bbR}{\mathbb{R}}
\newcommand{\bbC}{\mathbb{C}}
\newcommand{\bbU}{\mathbb{U}}

\begin{document}

\title{Error statistics and scalability of quantum error mitigation formulas}

\author{Dayue Qin}
\affiliation{Graduate School of China Academy of Engineering Physics, Beijing 100193, China}

\author{Yanzhu Chen}
\affiliation{Department of Physics, Virginia Tech, Blacksburg, Virginia 24061, USA}

\author{Ying Li}
\email{yli@gscaep.ac.cn}
\affiliation{Graduate School of China Academy of Engineering Physics, Beijing 100193, China}

\begin{abstract}
Quantum computing promises advantages over classical computing in many problems. Nevertheless, noise in quantum devices prevents most quantum algorithms from achieving the quantum advantage. Quantum error mitigation provides a variety of protocols to handle such noise using minimal qubit resources . While some of those protocols have been implemented in experiments for a few qubits, it remains unclear whether error mitigation will be effective in quantum circuits with tens to hundreds of qubits. In this paper, we apply statistics principles to quantum error mitigation and analyse the scaling behaviour of its intrinsic error. We find that the error increases linearly $O(\epsilon N)$ with the gate number $N$ before mitigation and sub-linearly $O(\epsilon' N^\gamma)$ after mitigation, where $\gamma \approx 0.5$, $\epsilon$ is the error rate of a quantum gate, and $\epsilon'$ is a protocol-dependent factor. The $\sqrt{N}$ scaling is a consequence of the law of large numbers, and it indicates that error mitigation can suppress the error by a larger factor in larger circuits. We propose the importance Clifford sampling as a key technique for error mitigation in large circuits to obtain this result.
\end{abstract}
\maketitle

\section{Introduction}

With the recent progress that quantum computers can have more than half a hundred qubits~\citep{arute_quantum_2019,gong_quantum_2021}, it is widely accepted that we are in the era of noisy intermediate-scale quantum (NISQ) technologies~\citep{preskill_quantum_2018}. A prominent feature of NISQ technologies is the potential for surpassing all classical computers in certain tasks, yet they cannot realize full quantum error correction and achieve fault tolerance due to noise and the limited number of physical qubits. Under the assumption of realistic noise models, the qubit overhead is thousands of physical qubits per logical qubit to reduce the chance of a logical error to the negligible level~\cite{fowler_surface_2012,ogorman_quantum_2017}. This requirement of quantum error correction is considerably beyond today's technologies.

Nevertheless, we can still perform computation tasks with NISQ devices. Protocols proposed recently allow us to bypass quantum error correction, which are termed quantum error mitigation~\cite{li_efficient_2017,temme_error_2017,endo_practical_2018,bonet-monroig_low-cost_2018,mcardle_error-mitigated_2019,mcclean_hybrid_2017,colless_computation_2018,huggins_virtual_2020,koczor_exponential_2021,kwon_hybrid_2020,smart_efficient_2020,endo_hybrid_2021}. Unlike error correction preserving the logical quantum state, error mitigation aims at recovering the error-free measurement outcome without physically preparing the error-free state. It can extract the correct computation result from a noisy device as long as the physical quantum state is not excessively damaged by the error accumulation~\cite{takagi_fundamental_2021}. For example, if the state becomes the maximally mixed state due to noise, there is nothing we can do to extract any useful information about the noise-free state. Recently, quantum algorithms using shallow circuits have been developed to minimise error accumulation. Quantum simulation algorithms based on variational, Lanczos and Monte Carlo methods are promising examples of such algorithms~\cite{peruzzo_variational_2014,mcclean_theory_2016,motta_determining_2020,yang_accelerated_2021,huggins_unbiasing_2021}. Although shallow-circuit algorithms and error mitigation protocols have been successful in proof-of-principle experiments~\cite{kandala_hardware-efficient_2017,arute_hartree-fock_2020,dumitrescu_cloud_2018,kandala_error_2019,song_quantum_2019,zhang_error-mitigated_2020,kim_scalable_2021,colless_computation_2018}, it remains unexplored how they will perform as we venture into the regime of useful applications, where the computation involves more than half a hundred qubits and the device noise permits error mitigation but not yet error correction.

In this work, we address how the computation error after mitigation scales with the circuit size. In many quantum algorithms, we use quantum circuits to evaluate the expected values of observables. For example, the Hamiltonian is evaluated in the variational quantum eigensolver~\cite{mcclean_theory_2016}. Because of noise, an actual quantum computer produces a biased expected value, and the bias usually increases with the circuit size due to the error accumulation. Among the error mitigation protocols, probabilistic error cancellation can completely remove the bias under ideal conditions~\cite{temme_error_2017,endo_practical_2018}. Under realistic conditions, however, all protocols leave a residual bias in the computation result. This residual bias depends on the protocol and circuit depth.

To draw a conclusion regardless of the protocol, we utilise a general formalism of error mitigation. In this formalism, we recover the observable in the error-free circuit using an error mitigation formula, which is a function of observables directly measured with noisy circuits. Many such formulas are inspired by our knowledge of quantum physics, such as error extrapolation~\cite{li_efficient_2017,temme_error_2017,giurgica-tiron_digital_2020,he_zero-noise_2020}, probabilistic error cancellation~\cite{temme_error_2017,endo_practical_2018} and virtual distillation~\cite{koczor_exponential_2021,huggins_virtual_2020,czarnik_qubit-efficient_2021,obrien_error_2021,huo_dual-state_2021}. Throughout this work, when a concrete error mitigation formula is needed for analysis, we take the three aforementioned protocols as examples. An alternative way to construct the formula is optimising a parameterised function with data of selected training circuits~\cite{strikis_learning-based_2021,czarnik_error_2020}. We find that the optimisation can suppress the scaling of the residual bias with respect to the circuit size.

For optimisation-based error mitigation protocols, we propose the importance Clifford sampling (ICS) as an efficient and scalable method to generate training circuits. Other than being practically useful in its own right, ICS lends us a tool to analyze the residual bias in the computation result. With its help, we show that the global depolarising model with circuit-dependent fluctuation is an effective phenomenological error model, which describes the impact of realistic error models. Using this  phenomenological model, we analyse the scaling behaviour of the residual bias. We find that the bias in the computation result after an optimised error mitigation process increases in proportion to $\sqrt{N}$, where $N$ is the gate number. In contrast, the bias is usually proportional to $N$ without error mitigation. Because error mitigation can suppress the error by a factor increasing with the circuit size, it is a feasible technique for large circuits. 

The Results section is organised as follows. After introducing the general formalism of error mitigation, we discuss the error scaling in the mitigation protocols using the global depolarising model, which will be validated subsequently as the effective phenomenological error model. Then we propose the ICS protocol, followed by a description of the important training circuits, the algorithms to generate them and an analysis of the sampling cost. We introduce the phenomenological error model and show that the fluctuation of the effective depolarising rate follows the $\sqrt{N}$ scaling, which is numerically verified. Finally, we show the same scaling relation between the bias and the gate number in error extrapolation, probabilistic error cancellation and virtual distillation. 

\section{Results}
\subsection{Error mitigation formula}
\label{sec:EMF}

\begin{figure}[tbp]
\begin{center}
\includegraphics[width=1\linewidth]{\figpath/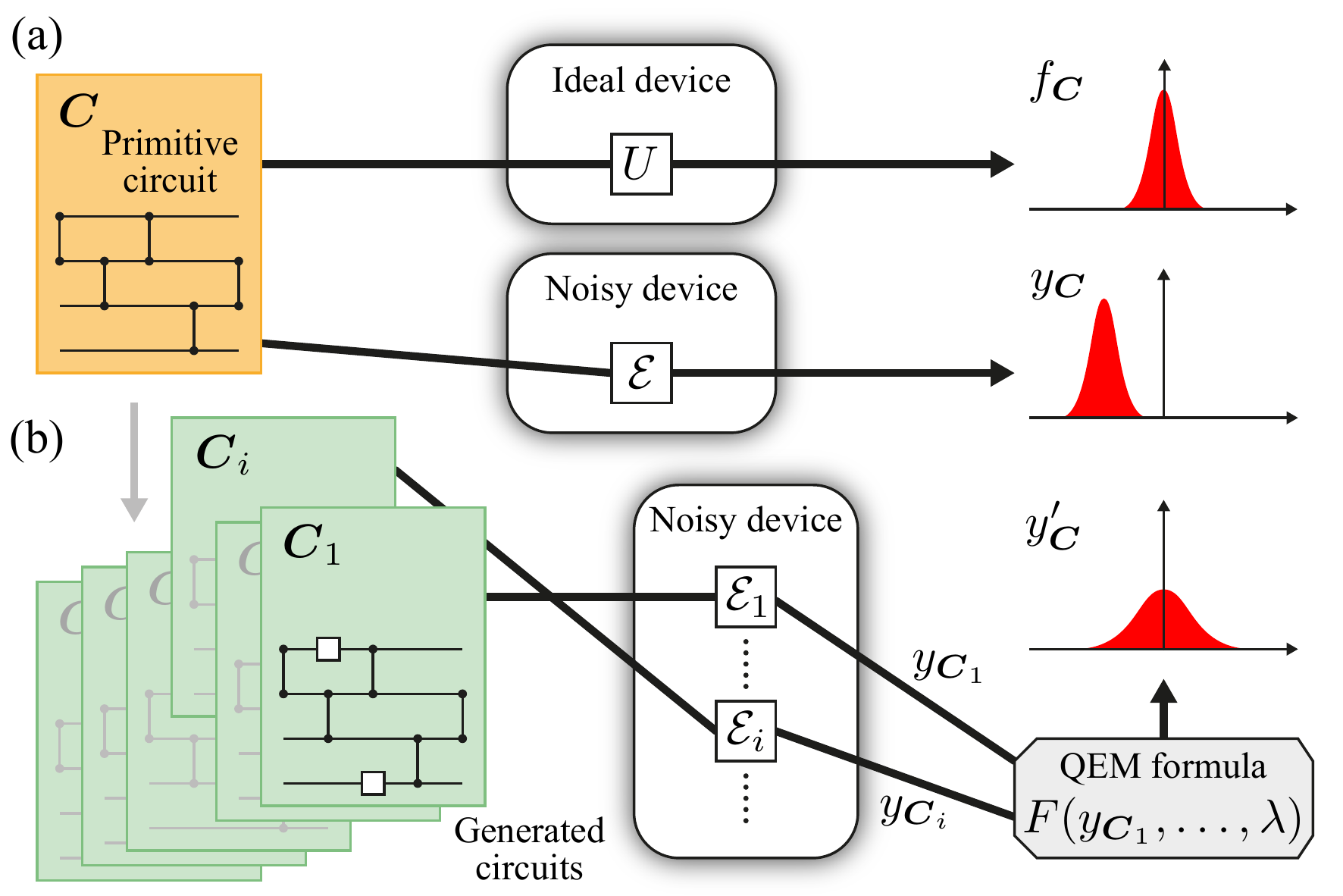}
\caption{
Schematic illustration of quantum error mitigation formulas. (a) Ideal and noisy quantum computing for the expected value of an observable. The distribution of the expected value is biased because of noise. (b) Error-mitigated quantum computing. The bias is corrected in quantum error mitigation (QEM).
}
\label{fig:QEM}
\end{center}
\end{figure}

First, we introduce the notations. In quantum computing, a quantum circuit consists of quantum gates. Let $U_j$ be the unitary operator of the $j$-th gate. The circuit with $N$ gates realises the transformation $U = U_N\cdots U_2 U_1$. Given the initial state of $n$ qubits $\ket{0}^{\otimes n}$ and observable $Q$, the expected value in the error-free circuit is $f_{\bfC} = \Tr[Q[U](\ketbra{0}{0}^{\otimes n})]$, where $[U](\bullet) = U\bullet U^\dag$. Here we use $\bfC = (U_1,\dots,U_N,Q)$ to denote the circuit with the observable specified. If the circuit is noisy, the transformation is inexact, and we use the completely-positive map $\mathcal{E}$ to denote the erroneous transformation. The expected value becomes $y_{\bfC} = \Tr[Q\mathcal{E}(\ketbra{0}{0}^{\otimes n})]$. Then, $y_{\bfC} - f_{\bfC}$ is the bias without error mitigation. Note that the error in the actual computing also depends on the statistical error due to finite measurement shots.

The general form of error mitigation formulas reads
\begin{eqnarray}
\label{eq:generalFormula}
y'_{\bfC} = F(y_{\bfC_1},y_{\bfC_2},\ldots,\lambda_1,\lambda_2,\ldots)
\end{eqnarray}
where $y'_{\bfC}$ is the result of the circuit $\bfC$ after error mitigation, $\bfC_1,\bfC_2,\dots$ are circuits generated from the primitive circuit $\bfC$, and $\lambda$'s denote parameters determined via error mitigation protocols. See Fig.~\ref{fig:QEM}. In quantum computing, we evaluate $y_{\bfC_i}$ using the noisy quantum computer and calculate the error-mitigated value $y'_{\bfC}$ according to the formula. The bias after error mitigation is $y'_{\bfC} - f_{\bfC}$. Next, we show how some specific error mitigation protocols fit into the general form.

Many error mitigation protocols have been proposed. See Ref.~\cite{endo_hybrid_2021} for a review. In this work, we take three protocols as examples: error extrapolation, probabilistic error cancellation and virtual distillation. These protocols are applicable to any quantum algorithm evaluating expected values and can largely reduce the error. We give a minimal description here and leave a more detailed overview to Appendix~\ref{sec:appEM}.

In error extrapolation using a polynomial fitting function~\cite{temme_error_2017,giurgica-tiron_digital_2020} , the error mitigation formula is
\begin{eqnarray}
y'_{\bfC} = \sum_{i} q_i y_{\bfC_i},
\label{eq:EMFL}
\end{eqnarray}
where $\bfC_i$ is the primitive circuit with noise increased by a factor of $r_i$, and coefficients $q_i$ are determined by noise amplification factors (i.e.~$r_i$). For example, for the linear extrapolation with $r_1 = 1$ and $r_2 = 2$, the formula is
\begin{eqnarray}
y'_{\bfC} = 2y_{\bfC_1} - y_{\bfC_2}.
\label{eq:LEF}
\end{eqnarray}

In probabilistic error cancellation, the completely-positive map of the error-free circuit is expressed as a linear combination of erroneous maps, i.e.~
\begin{eqnarray}
[U] = \sum_i q_i \mathcal{E}_i,
\label{eq:PEC}
\end{eqnarray}
where $q_i$ are quasi-probabilities, and $\mathcal{E}_i$ is the map of a noisy circuit $\bfC_i$. Here $\bfC_i$ is generated by, for example, replacing or adding some gates in the primitive circuit $\bfC$. We can work out the quasi-probability decomposition with gate set tomography data~\cite{endo_practical_2018} or in a learning manner~\cite{strikis_learning-based_2021}. Given the decomposition, the error mitigation formula is the same as Eq.~(\ref{eq:EMFL}), but coefficients and circuits are different from error extrapolation.

In virtual distillation, $k$ copies of the erroneous state $\rho$ are used to evaluate the observable in a distilled state without physically preparing it. Given the primitive circuit $\bfC$ that prepares the state $\rho$, the circuit $\bfC_1$ is to evaluate $y_{\bfC_1} = \Tr(Q\rho^k)$, and the circuit $\bfC_2$ is to evaluate $y_{\bfC_2} = \Tr(\rho^k)$. Then the error mitigation formula reads
\begin{eqnarray}
y'_{\bfC} = \frac{y_{\bfC_1}}{y_{\bfC_2}}.
\end{eqnarray}
It is similar in related protocols, e.g.~verified phase estimation~\cite{obrien_error_2021} and dual-state purification~\cite{huo_dual-state_2021}.

\subsection{Bias in the global depolarising model}
\label{sec:GDM}

\begin{figure}[tbp]
\begin{center}
\includegraphics[width=1\linewidth]{\figpath/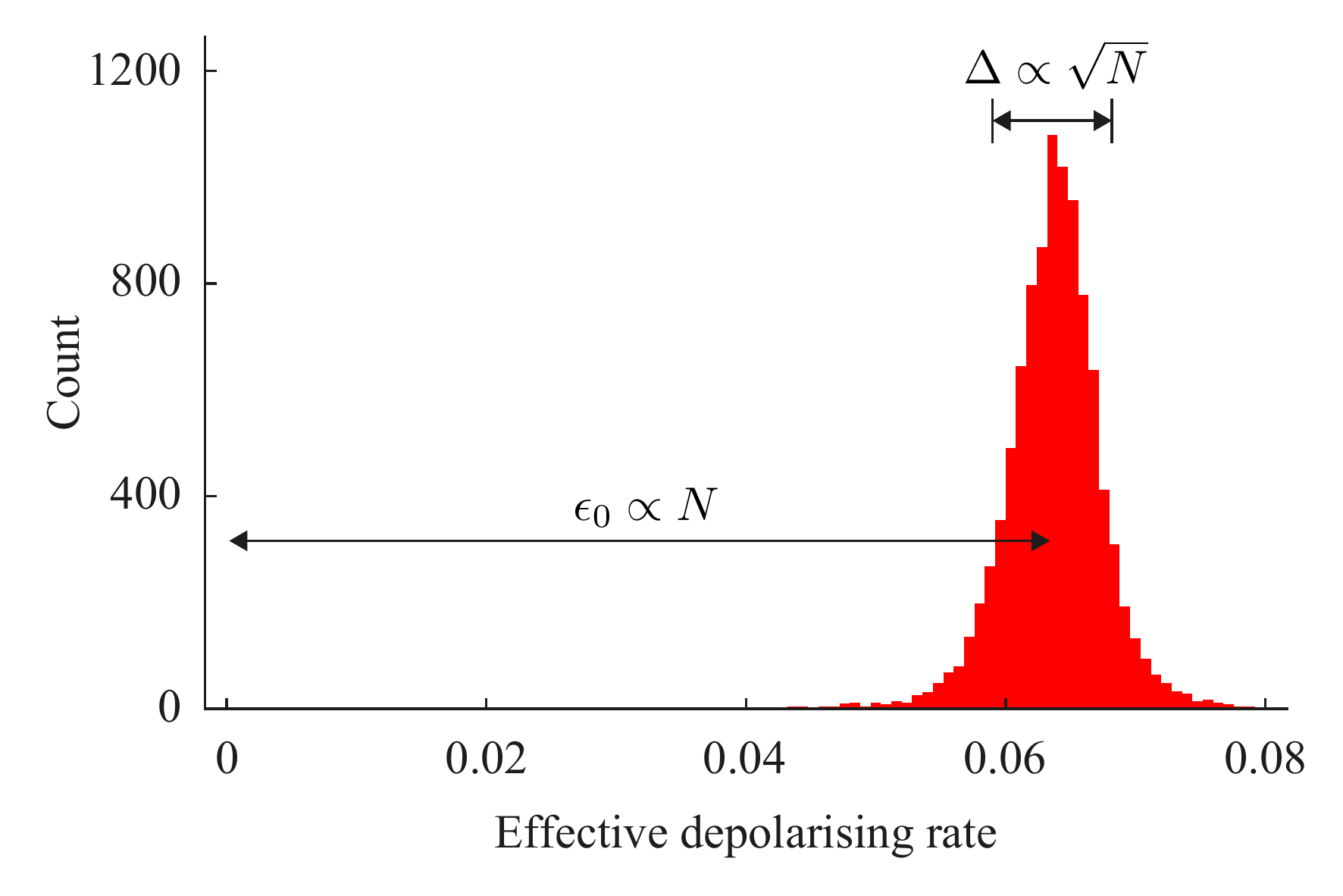}
\caption{
Distribution of the effective depolarising rate in the phenomenological error model. In the model, the impact of errors in a noisy circuit is characterised by the global depolarising model with the circuit-dependent depolarising rate $\epsilon_{\bfC}$. The histogram is generated using six-qubit periodic-cycling circuits with $72$ two-qubit gates under the gate depolarising noise. The error rate per gate is $0.001$. Single-qubit gates are randomly sampled from the set of single-qubit unitaries with the weight $f_{\bfC}^2$. The average depolarising rate is proportional to the gate number $N$, and the standard deviation is proportional to $\sqrt{N}$.
}
\label{fig:model}
\end{center}
\end{figure}

Before considering realistic error models, we take the global depolarising model as an example to discuss the bias in error mitigation formulas. In this section, we show that, if the error mitigation protocols are perfectly implemented, probabilistic error cancellation and learning-based error mitigation can reduce the bias to zero, while linear extrapolation and virtual distillation with two copies can reduce the bias from $O(N\epsilon)$ to $O(N^2\epsilon^2)$, where $N$ is the gate number and $\epsilon$ is the depolarising rate per gate. In the section of ``Phenomenological error model'' we will show that the global depolarising model successfully captures the influence of realistic noise and can be used as a phenomenological model.

In the global depolarising model, the $j$-th gate with error is described by the map $\mathcal{G}_j = (1-\epsilon)[U_j] + \epsilon \mathcal{D}$ acting on the whole input state, where $\epsilon$ is the gate depolarising rate, $\mathcal{D}(\bullet) = \Tr(\bullet) \rho_m$ is the depolarising map, and $\rho_m = \openone/2^n$ is the maximally mixed state. Without loss of generality, we assume that the observable is a traceless operator, and we have $y_{\bfC} = (1-\epsilon)^N f_{\bfC} = f_{\bfC} + O(\epsilon N)$. The bias increases linearly with the gate number when $N$ is significantly smaller than $\epsilon^{-1}$. In the limit of large $N$, the bias approaches a finite value if the observable is bounded.

We take linear extrapolation as an example of error extrapolation. We can construct two noisy circuits using original gates and double-noise gates, respectively. Let $\mathcal{G}_j' = (1-2\epsilon)[U_j] + 2\epsilon \mathcal{D}$ be the gate with the doubled depolarising rate, two circuits labelled by $i=1, 2$ produce expected values $y_{\bfC_i} = \Tr[Q\mathcal{E}_i(\ketbra{0}{0}^{\otimes n})]$, where $\mathcal{E}_1 = \mathcal{G}_N \cdots \mathcal{G}_2 \mathcal{G}_1$ and $\mathcal{E}_2 = \mathcal{G}_N' \cdots \mathcal{G}_2' \mathcal{G}_1'$. Then, Eq.~(\ref{eq:LEF}) leads to the error-mitigated expected value
\begin{eqnarray}
y'_{\bfC} &=& 2(1-\epsilon)^N f_{\bfC} - (1-2\epsilon)^N f_{\bfC} \notag \\
&=& f_{\bfC} + O(\epsilon^2 N^2).
\end{eqnarray}
We can find that the bias in the linear extrapolation formula increases quadratically with the gate number because the linear extrapolation eliminates the first-order contribution of errors.

In probabilistic error cancellation, we take the quasi-probability decomposition of each gate as
\begin{eqnarray}
[U_j] &=& \frac{1}{1-\epsilon} \mathcal{G}_j - \frac{\epsilon}{1-\epsilon} \mathcal{D}.
\end{eqnarray}
This decomposition means that we can correct the error by stochastically replacing the original gate $\mathcal{G}_j$ with the depolarising map $\mathcal{D}$ according to a quasi-probability distribution. The decomposition formula of the entire circuit reads
\begin{eqnarray}
[U] &=& \prod_{j=1}^N \left[ \frac{1}{1-\epsilon} \mathcal{G}_j - \frac{\epsilon}{1-\epsilon} \mathcal{D} \right] \notag \\
&=& \frac{1}{(1-\epsilon)^N} \mathcal{E}_1 - \frac{\epsilon}{(1-\epsilon)^N} \mathcal{E}_2 + \cdots,
\end{eqnarray}
where $\mathcal{E}_1 = \mathcal{G}_N \cdots \mathcal{G}_2 \mathcal{G}_1$ corresponding to the primitive circuit, $\mathcal{E}_2 = \mathcal{G}_N \cdots \mathcal{G}_2 \mathcal{D}$ in which the first gate is replaced, and so on. Then the error mitigation formula is
\begin{eqnarray}
y'_{\bfC} = \frac{1}{(1-\epsilon)^N} y_{\bfC_1} - \frac{\epsilon}{(1-\epsilon)^N} y_{\bfC_2} + \cdots = f_{\bfC}.
\label{eq:EMFec}
\end{eqnarray}
Here, we have used that $y_{\bfC_i} = 0$ if any gate is replaced with $\mathcal{D}$. Therefore, the residual bias is zero.

Lastly, we consider virtual distillation. The final state of $N$ gates with the depolarising error is
\begin{eqnarray}
\rho &=& (1-\epsilon_t) U\ketbra{0}{0}^{\otimes n} U^\dag + \epsilon_t \rho_m,
\end{eqnarray}
where $\epsilon_t = 1-(1-\epsilon)^N$. Take the second-order virtual distillation (i.e.~$k = 2$) as an example, the error-mitigated expected value is
\begin{eqnarray}
y'_{\bfC} &=& \frac{(1-\epsilon_t)^2 + 2^{1-n}(1-\epsilon_t)\epsilon_t}{(1-\epsilon_t)^2 + 2^{1-n}(1-\epsilon_t)\epsilon_t + 2^{- n}\epsilon_t^2}f_{\bfC} \notag \\
&=& f_{\bfC} + O(\epsilon^2 N^2).
\label{eq:EMFvd}
\end{eqnarray}
Therefore, the bias in the second-order virtual distillation increases quadratically with the gate number, which is the natural consequence of the second-order distillation formalism.

So far we have been considering ideal conditions. Under realistic conditions, imperfections in the implementation cause an additional contribution to the bias. For example, zero-bias probabilistic error cancellation requires exact knowledge about the depolarising rate. If the depolarising rate is thought to be $\epsilon'$ instead of its actual value $\epsilon$ and we work out the error mitigation formula with $\epsilon'$, we have $y'_{\bfC} = (1-\epsilon)^N/(1-\epsilon')^N f_{\bfC}$. Then, the bias of the error mitigation formula is $O((\epsilon''-\epsilon)N)$, which is finite and increases linearly with the gate number. It is similar for error extrapolation, in which the bias scales linearly if the noise is not increased exactly as designed.

Next, we analyse the bias in learning-based error mitigation. The optimisation of an ansatz function is a flexible approach for working out a proper error mitigation formula. Various ansatz functions have been proposed~\cite{strikis_learning-based_2021,czarnik_error_2020,bultrini_unifying_2021}. In this work, we consider a general framework of this approach and focus on the scaling of the bias with respect to the gate number.

One way to compose an ansatz function is by modifying a specific-form formula. Taking the linear error extrapolation as an example, we parameterise the formula as
\begin{eqnarray}
y'_{\bfC} = \lambda y_{\bfC_1} + (1-\lambda) y_{\bfC_2}.
\label{eq:EMFlambda}
\end{eqnarray}
We determine $\lambda$ by minimising the bias for a set of circuits, which are called training circuits. To evaluate the bias, the error-free expected value must be known. This condition limits the choice of training circuits. We can use only one training circuit $\bfT$ and the corresponding data $(y_{\bfT_1},y_{\bfT_2},f_{\bfT})$ to determine $\lambda$ for the ansatz considered here. The bias of the training circuit is minimised at
\begin{eqnarray}
\lambda^* = \frac{f_{\bfT} - y_{\bfT_2}}{y_{\bfT_1} - y_{\bfT_2}}.
\label{eq:lambda}
\end{eqnarray}
For the global depolarising model, the optimal parameter is $\lambda^* = [1-(1-2\epsilon)^N]/[(1-\epsilon)^N-(1-2\epsilon)^N]$. If we take $\lambda = \lambda^*$ in the error mitigation formula, the bias is zero for all circuits with the same gate number $N$. Therefore, the linear error extrapolation becomes bias-free after the optimisation.

It is similar for other error mitigation protocols. For probabilistic error cancellation, we can take the depolarising rate $\epsilon$ in Eq.~(\ref{eq:EMFec}) as the variational parameter, assuming the actual depolarising rate is unknown. We can find the optimal value of $\epsilon$ with data of a training circuit, and the optimal value must be the actual depolarising rate. Then, the error mitigation formula taking the optimal parameter is bias-free for all circuits. For virtual distillation, we can choose the ansatz $y_{\bfC}' = \lambda \frac{y_{\bfC_1}}{y_{\bfC_2}}$. According to Eq.~(\ref{eq:EMFvd}), the bias is zero when $\lambda$ cancels the factor before $f_{\bfC}$.

We have seen that the learning-based approach can reduce the bias in error mitigation. According to the global depolarising model, the bias is zero in all examples. We get this perfect result because the global depolarising model is free of fluctuation, i.e.~errors of all gates have the same impact on the expected value. The impact is a factor of $1-\epsilon$. Without the fluctuation, there are many simple error mitigation formulas that can simultaneously and completely correct the bias for all circuits.

In error models with fluctuation, the optimised error mitigation formula has a finite bias, and the bias increases with the gate number. Usually, errors are localised in many actual quantum computing systems, e.g.~superconducting qubits and trapped ions. The error associated with a gate only affects qubits at the location of the gate (rather than the entire quantum register as in the global depolarising model). The contribution of an error to the bias depends on its location and the circuit. For example, if the observable is the Pauli operator $X$ of qubit-1, errors localised on qubit-2 do not affect the observable; A phase-flip error before the measurement changes the sign of $X$ but preserves the sign if we modify the circuit by inserting a Hadamard gate before the measurement. The fluctuation of error contributions causes a finite bias, i.e.~the error mitigation formula cannot simultaneously compensate for all errors for all circuits. Assuming we can successfully compensate for the average contribution of errors, the residual bias is due to the fluctuation across different circuits. We find that in a large class of error mitigation formulas, the fluctuation-caused bias is proportional to $\sqrt{N}$. Later, we will show that the global depolarising model with fluctuation is an effective phenomenological model to characterise the impact of errors in realistic error models, see Fig.~\ref{fig:model}. 

\subsection{Importance Clifford sampling}
\label{sec:ICS}

In this section, we address the question of how to efficiently sample large training circuits by proposing sampling algorithms whose resource costs scale linearly with the circuit size. These training circuits are Clifford circuits sharing the same circuit frame as the original noisy circuit, for which the ideal measurements take non-zero expected values.

A classical computer can efficiently simulate Clifford circuits, in which all gates are Clifford gates. Because the error-free expected value $f_{\bfC}$ of a Clifford circuit is computable~\cite{aaronson_improved_2004,anders_fast_2006}, we can take them as training circuits. However, not every Clifford circuit is suitable. We take Eq.~(\ref{eq:lambda}) as an example. If the training circuit $\bfT$ has a zero expected value, i.e.~$f_{\bfT} = 0$, erroneous expected values are all zero, i.e.~$y_{\bfT_1} = y_{\bfT_2} = 0$. In this case, we cannot use the equation to determine the optimal parameter. Therefore, to find the optimal parameter, we need a training circuit $\bfT$ whose expected value is nonzero.

It is general that some training circuits are more important than others in the learning-based approach. To optimise the error mitigation formula, we need a measure of its overall performance in various circuits. We take the mean squared error (MSE) as an example, which reads
\begin{eqnarray}
L_{\bbR} = \mean{(y_{\bfC} - f_{\bfC})^2}_{\bbR},
\end{eqnarray}
where $\mean{g(\bfC)}_{\bbR} \equiv \frac{1}{\abs{\bbR}} \sum_{\bfC\in\bbR} g(\bfC)$ is the average of the real-valued circuit function $g(\bfC)$ over the circuit set $\bbR$. Importance sampling is a crucial technique in statistics, in which the probability of a sample is proportional to the magnitude of its value, i.e.~$(y_{\bfC} - f_{\bfC})^2$ in MSE. According to importance sampling, we prefer training circuits with a larger bias over those with a smaller bias. The larger-bias circuits, i.e.~error-sensitive circuits, can provide more information about noise in the circuit.

The question of sampling training circuits has two parts. The first part is how to efficiently generate an error-sensitive circuit. The second part is how to draw samples according to a distribution. We address the first part in the ``Circuit generation'' section and the second part in the ``Circuit frame'' and ``Sampling algorithms'' sections.

\subsection{Circuit generation}
\label{sec:generation}

\begin{figure}[tbp]
\begin{center}
\includegraphics[width=1\linewidth]{\figpath/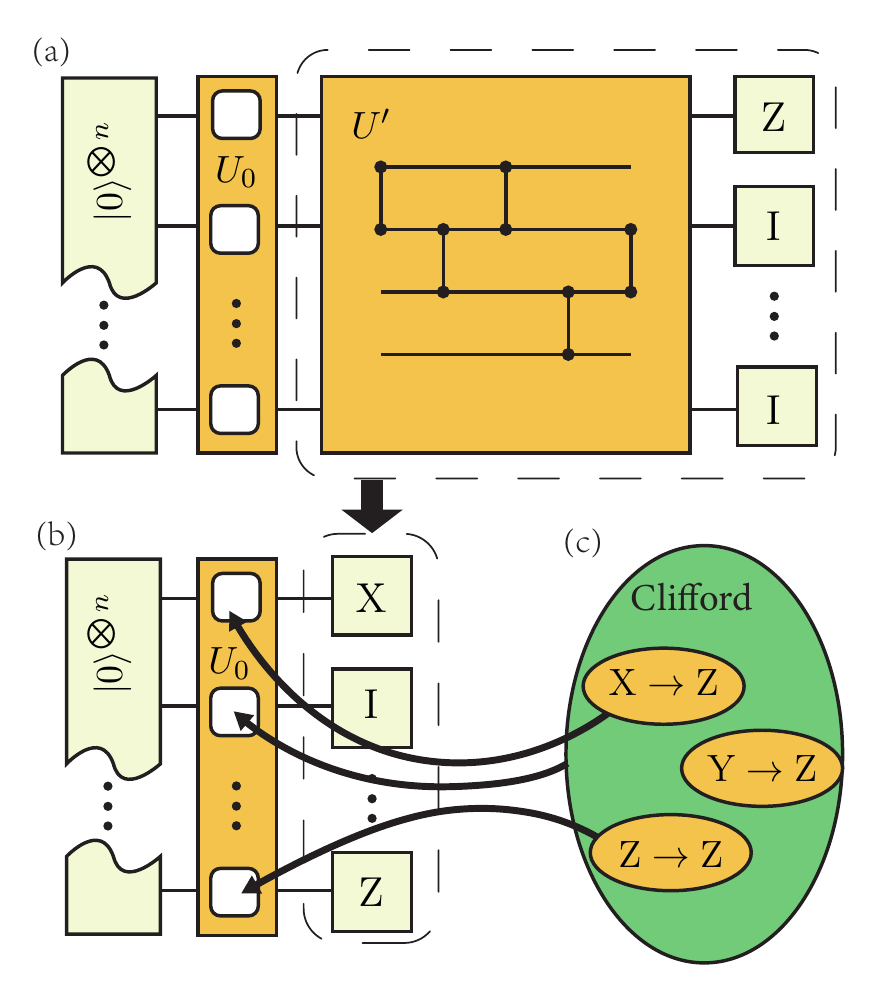}
\caption{
Error-sensitive circuit generation. We compose an error-sensitive circuit with two sections $U_0$ and $U'$, as shown in (a). $U'$ is a Clifford operator. The observable is a Pauli operator , e.g.~$Q = Z\otimes I\otimes\cdots\otimes I$. $U'$ and the observable is equivalent to an effective observable $Q_{U'} = X\otimes I\otimes\cdots\otimes Z$, as shown in (b). Gates in $U_0$ are taken from the group of single-qubit Clifford gates. We choose the gates such that all non-identity Pauli operators in $Q_{U'}$ are mapped to $\pm Z$, as shown in (c).
}
\label{fig:SAM_ES}
\end{center}
\end{figure}

There are different approaches of generating an error-sensitive circuit. For example, we can randomly select a circuit and calculate the expected value, and we take it as a training circuit only if the expected value is nonzero. This approach works only when the circuit size is small because circuits with a nonzero expected value are rare in large Clifford circuits. An approach usually used in randomised benchmarking is reversing the transformation by adding an additional unitary at the end of the circuit~\cite{magesan_scalable_2011}. We will not take this approach because the additional unitary may significantly increase the total gate number in multi-qubit circuits. We want to generate training circuits with a specific gate number, such that the error mitigation formula is optimised for circuits with the same gate number.

In the following, we focus on the case that the observable $Q$ is a Pauli operator. In the standard model of quantum computing, qubits at the end of the circuit are measured in the computation basis, i.e.~the Pauli operator $Z$ is measured. One can adjust the measurement basis by inserting gates before the measurement. For example, by inserting single-qubit Clifford gates before the measurement, we can measure any Pauli operator. For a general observable, a way to evaluate its expected value is by expressing it as a linear combination of Pauli operators and computing the expected value of each term.

The expected value of a Pauli operator in a Clifford circuit takes three values $0$ and $\pm 1$. We can reexpress the error-free expected value as $f_{\bfC} = \Tr(Q_U\ketbra{0}{0}^{\otimes n})$, where $Q_U = U^\dag QU$ is the effective observable. When $U$ is Clifford, $Q_U$ is a Pauli operator. Let $P_i = I,X,Y,Z$ be the single-qubit Pauli operator on qubit-$i$, $Q_U = \pm P_1\otimes P_2\otimes\cdots\otimes P_n$. Then $f_{\bfC} = \pm \prod_{i=1}^n \bra{0}P_i\ket{0}$. If any single-qubit Pauli operator $P_i$ is $X$ or $Y$, the expected value is zero. If all $P_i$ are $I$ or $Z$, $f_{\bfC} = \pm 1$, and the sign is the same as $Q_U$. For a randomly generated Clifford circuit, it is likely that some single-qubit Pauli operators contained in $Q_U$ are $X$ or $Y$, i.e.~$f_{\bfC} = 0$.

We can deterministically generate an error-sensitive circuit as follows. The setup is shown in Fig.~\ref{fig:SAM_ES}. The overall unitary transformation of the circuit is $U = U'U_0$, where $U_0 = R_1\otimes R_2\otimes\cdots\otimes R_n$ is one layer of single-qubit gates, and $R_i$ is the gate on qubit-$i$. First, given the gate number, we generate a random Clifford circuit, which realises the unitary $U'$. If $U_0 = \openone$, the effective observable is $Q_{U'} = \pm P_1'\otimes P_2'\otimes\cdots\otimes P_n'$. Given $Q$ and $U'$, we can efficiently work out this expression of $Q_{U'}$ on a classical computer. Second, we determine single-qubit gates in $U_0$: we take a Clifford $R_i$ satisfying $R_i^\dag P_i' R_i = \pm Z,I$. For the final circuit $U = U'U_0$, single-qubit Pauli operators in its effective observable $Q_U$ are either $I$ or $Z$. Then, the expected value is $f_{\bfC} = \pm 1$.

\subsection{Circuit frame}
\label{sec:frame}

\begin{figure}[tbp]
\begin{center}
\includegraphics[width=1\linewidth]{\figpath/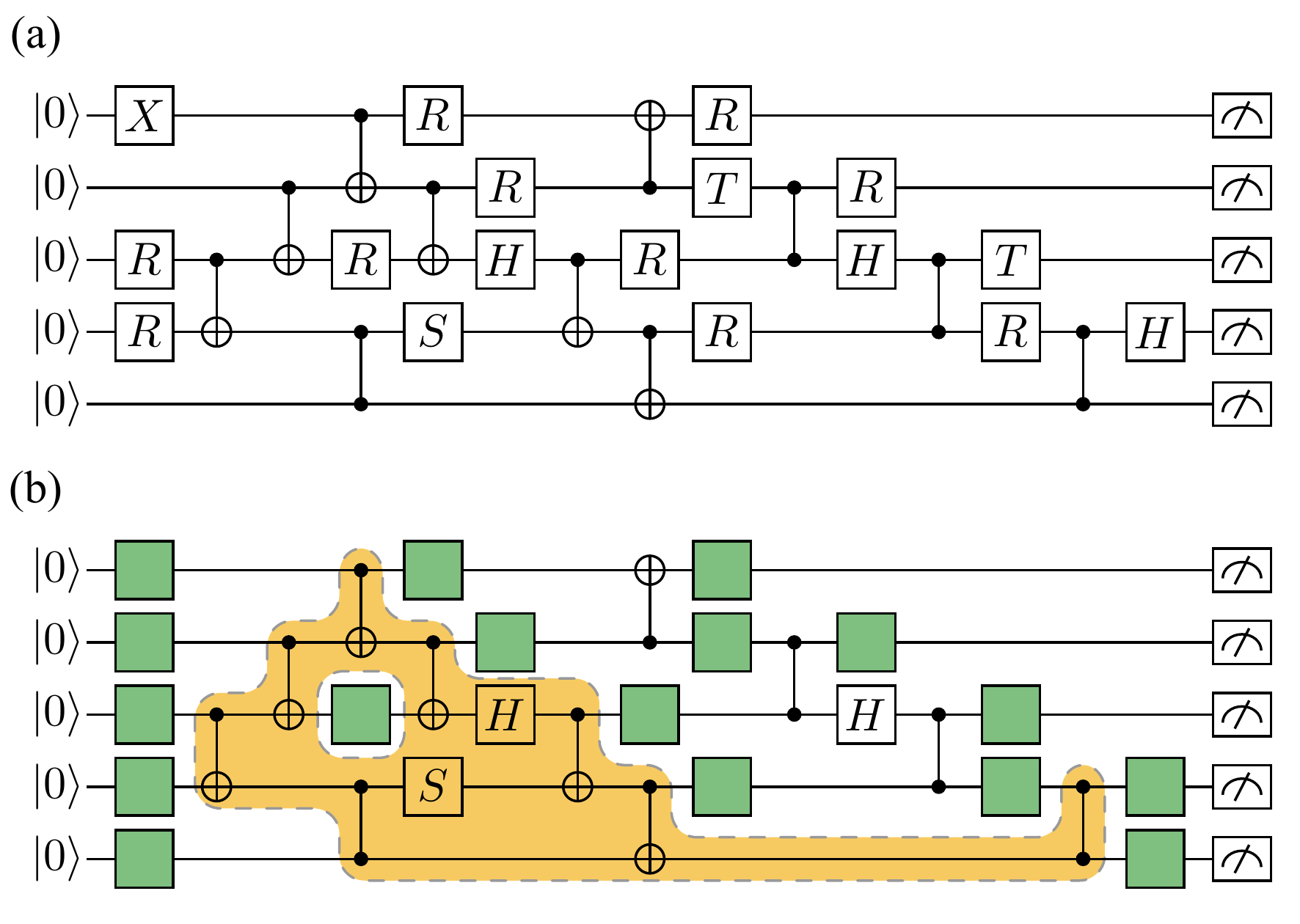}
\caption{
Quantum circuit and circuit frame. (a) The circuit for a specific task. Single-qubit gates $X$, $H$ and $S$ are Clifford, and gates $T$ and $R$ are non-Clifford. (b) The task-dependent circuit frame. Green boxes are slots for variable single-qubit gates. Clifford gates in the yellow region with dashed borders form a composite Clifford gate.
}
\label{fig:circuit_frame}
\end{center}
\end{figure}

In the learning-based error mitigation, we aim at an optimised error mitigation formula that works for a set of circuits, including training circuits and circuits useful in some computation tasks. Choosing the target circuit set is important. When the circuit set is larger, it is harder to find a formula suitable for every circuit. Therefore, we want to be focusing on a circuit set relevant to some tasks to minimise bias. A way to construct a task-relevant circuit set is by taking circuits with the same pattern of multi-qubit Clifford gates, see Fig.~\ref{fig:circuit_frame}. This pattern is called the circuit frame. In many quantum computing systems, such as superconducting qubits and trapped ions, the error rates of single-qubit gates are much lower than multi-qubit gates. Errors occurring in a circuit are mainly determined by multi-qubit gates. Therefore, all the circuits with the same frame have approximately the same errors, and we are able to correct them using the same error mitigation formula.

In the fixed-frame circuit set, single-qubit gates are variables. As shown in Fig.~\ref{fig:circuit_frame}, the frame includes the qubit initialisation, multi-qubit Clifford gates and measurement. Fixing these operations, we change single-qubit gates to generate the circuit set. We call each variable single-qubit gate a slot. In Ref.~\cite{strikis_learning-based_2021}, a setup with slots after each multi-qubit gate is proposed. Here we reduce the slot number to minimise the circuit set. We only take locations of single-qubit non-Clifford gates in the task circuit as slots and add two layers of slots after the initialisation and before the measurement, respectively. The reason is that a sequence of Clifford gates not interrupted by any non-Clifford gate can be treated as one multi-qubit Clifford gate.

The minimised slots have sufficient degrees of freedom for implementing Pauli twirling and probabilistic error cancellation for general error models. A Pauli error is an unwanted Pauli transformation stochastically occurring in the circuit. In Pauli twirling, we convert general errors into Pauli errors by randomly applying Pauli gates before and after each Clifford gate. We can correct a Pauli error by applying a Pauli gate to undo the error. Relevant discussions can be found in Ref.~\cite{strikis_learning-based_2021}.

With the frame determined, a circuit depends on the choice of single-qubit gates. Let $\bfC = (U_1,\dots,U_N,Q)$ be a circuit (with two layers of single-qubit gates after the initialisation and before the measurement, respectively). The corresponding frame is $F = (\dots,U_i,\bullet_k,\ldots,U_j,\bullet_q,\ldots,Q)$, where $U_i$ is a gate on the frame, and $\bullet_k$ denotes a slot on qubit-$k$. In other words, $F$ is the same as $\bfC$ except that gates in slots are replaced with $\bullet_k$. Formally, if $S = \{i_1,i_2,\dots\}$ are labels of slots and $K = \{k_{i_1},k_{i_2},\dots\}$ are corresponding qubits, the frame is $F = (F_1,\dots,F_N,Q)$, where $F_i = U_i$ if $i\notin S$, and $F_i = \bullet_{k_i}$ if $i\in S$. Then, we can reexpress the circuit as $\bfC = [F,R_1,R_2,\ldots]$, where $R_l$ is the single-qubit gate in the $l$-th slot, i.e.~$U_{i_l} = I^{\otimes (k_{i_l}-1)}\otimes R_l\otimes I^{\otimes (n-k_{i_l})}$.

To generate training circuits of the fixed frame, we can randomly draw the gate on each slot from the $24$ single-qubit Clifford gates. Because the frame is formed of Clifford gates, the entire circuit constructed in this way is Clifford. It is likely that such a random circuit has a zero expected value. We can work out a circuit with a non-zero expected value by adjusting the first-layer gates, i.e.~gates after the initialisation, as described in in the previous section. We give details of this procedure in Algorithm~\ref{alg:generation}.

\begin{figure}
\begin{minipage}{\linewidth}
\begin{algorithm}[H]
{\small
\begin{algorithmic}[1]
\caption{{\small Generation of error-sensitive circuits.}}
\label{alg:generation}
\Statex
\Function{EScircuit}{$F,\bar{R}$}
\State Compose the candidate circuit $\bfC' = [F,I,\ldots,I,R_{n+1},\ldots,R_{N_R}]$.
\State Calculate $Q_{U'} = U^{\prime\dag} QU'$.
\State Calculate $(P_1',P_2',\ldots,P_n')$ according to $Q_{U'} = \pm P_1'\otimes P_2'\otimes\cdots\otimes P_n'$.
\For{$i=1$ to $n$}
\Repeat
\State Choose a random $R_i $ from $C_1$.
\Until{$R_i^\dag P_i' R_i = \pm Z,I$}
\EndFor
\State Compose the error-sensitive circuit $\bfC = [F,R_1,\ldots,R_n,R_{n+1},\ldots,R_{N_R}]$.
\State Output $\bfC$.
\EndFunction
\end{algorithmic}
}
\end{algorithm}
\end{minipage}
\end{figure}

\subsection{Sampling algorithms}
\label{sec:algorithms}

We give two algorithms for sampling error-sensitive Clifford circuits in Algorithms~\ref{alg:nonU}~and~\ref{alg:U}. For clarity, we use the following notations in the algorithms. $F$ is the circuit frame, $Q$ is the observable, $n$ is the qubit number, $N_R$ is the slot number, and $N_T$ is the sample number. $C_1$ is the single-qubit Clifford group with $24$ elements. $U = U_N \cdots U_2 U_1$ is the unitary transformation of the circuit $\bfC = (U_1,\dots,U_N,Q) = [F,R_1,R_2,\ldots]$. We use $\bar{R} = (R_{n+1},R_{n+2},\dots,R_{N_R})$ to denote an ordered set of single-qubit Clifford gates, and $R_1,R_2,\ldots,R_n$ are gates in the first-layer slots. $w(\bfC)$ is the weight of the Clifford circuit $\bfC$: $Q_U = U^\dag QU = \pm P_1\otimes P_2\otimes\cdots\otimes P_n$ is a tensor product of Pauli operators, then $w(\bfC)$ is the number of non-identity Pauli operators in the product, i.e.~
\begin{eqnarray}
w(\bfC) \equiv n - \sum_{i=1}^n \delta_{I,P_i},
\end{eqnarray}
where $\delta_{I,P_i}=1$ if $P_i=I$, and $\delta_{I,P_i}=0$ otherwise.
In Algorithm~\ref{alg:U}, we employ the Metropolis-Hasting algorithm to realise a uniform distribution of error-sensitive circuits, which requires a conditional distribution $g(\bar{R}' \vert \bar{R})$ for suggesting a candidate sample. For example, we can take the conditional distribution as follows: we update gates in some randomly selected slots with newly generated random gates and keep gates in other slots unchanged.

\begin{figure}
\begin{minipage}{\linewidth}
\begin{algorithm}[H]
{\small
\begin{algorithmic}[1]
\caption{{\small Non-uniform importance Clifford sampling.}}
\label{alg:nonU}
\Statex
\State Input $F$.
\For{$t=1$ to $N_T$}
\For{$i=n+1$ to $N_R$}
\State Choose a random $R_i$ from $C_1$.
\EndFor
\State Call \Call{EScircuit}{$F,\bar{R}$} to generate $\bfC$.
\State Output $\bfC_t = \bfC$.
\EndFor
\end{algorithmic}
}
\end{algorithm}
\end{minipage}
\end{figure}

\begin{figure}
\begin{minipage}{\linewidth}
\begin{algorithm}[H]
{\small
\begin{algorithmic}[1]
\caption{{\small Uniform importance Clifford sampling.}}
\label{alg:U}
\Statex
\State Input $F$, a conditional distribution $g(\bar{R}' \vert \bar{R})$ and an initial slot-gate pattern $\bar{R}^{(0)}$.
\State Set $t = 0$.
\State Call \Call{EScircuit}{$F,\bar{R}^{(0)}$} to generate $\bfC$.
\State Take $\bfC_0 = \bfC$.
\For{$t=1$ to $N_T$}
\State Generate a random candidate of slot-gate pattern $\bar{R}^{(t)}$ according to $g(\bar{R}^{(t)}\vert \bar{R}^{(t-1)})$.
\State Call \Call{EScircuit}{$F,\bar{R}^{(t)}$} to generate $\bfC$.
\State Calculate the acceptance probability
$$A = \min\left(1,\frac{3^{-w(\bfC)}}{3^{-w(\bfC_{t-1})}} \frac{g(\bar{R}^{(t-1)}\vert \bar{R}^{(t)})}{g(\bar{R}^{(t)}\vert \bar{R}^{(t-1)})}\right).$$
\State Generate a uniform random number $u\in [0,1]$.
\State Accept and set $\bfC_t = \bfC$ if $u\leq A$.
\State Reject and set $\bfC_t = \bfC_{t-1}$ if $u> A$.
\State Output $\bfC_t$.
\EndFor
\end{algorithmic}
}
\end{algorithm}
\end{minipage}
\end{figure}

There is a relation between Clifford sampling and unitary sampling which allows us to estimate the bias distribution in general unitary circuits using Clifford circuits. We use $\bbC$ to denote the set of Clifford circuits and $\bbU$ to denote the set of all unitary circuits with the same frame. For a frame with $N_R$ slots, the total number of Clifford circuits is $\abs{\bbC} = 24^{N_R}$, i.e.~each slot takes one of $24$ single-qubit Clifford gates. In $\bbU$, each slot can take any single-qubit unitary. When errors are independent of the choice of single-qubit gates, MSEs are the same for the two circuit sets, i.e.~$L_{\bbU} = L_{\bbC}$~\cite{wang_scalable_2021}. Because the set $\bbC$ is large, we need to use the Monte Carlo method to evaluate $L_{\bbC}$.

There is a similar relation between ICS and unitary sampling. Error-sensitive circuits are a subset of all Clifford circuits, denoted by $\bbC^{ES}$. According to Algorithm~\ref{alg:generation}, given slot gates $\bar{R} = (R_{n+1},R_{n+2},\dots,R_{N_R})$, the number of error-sensitive circuits is $8^{w(\bfC)}24^{n-w(\bfC)}$. If $P_i' = I$, $R_i^\dag P_i' R_i = I$ for all $24$ single-qubit Clifford gates, which contributes a factor of $24$; If $P_i' \neq I$, $R_i^\dag P_i' R_i = \pm Z$ for $8$ single-qubit Clifford gates, which contributes a factor of $8$. The number of different $\bar{R}$'s is $24^{N_R-n}$, then the total number of error-sensitive circuits is
\begin{eqnarray}
\abs{\bbC^{ES}} = \sum_{j=1}^{24^{N_R-n}} 8^{w(\bfC_j)}24^{n-w(\bfC_j)},
\end{eqnarray}
where $\bfC_j$ are circuits with different $\bar{R}$'s. In a Clifford circuit, a Pauli error either preserves the Pauli observable or flips its sign. As a result, non-sensitive Clifford circuits do not respond to Pauli errors, i.e~$y_{\bfC} = f_{\bfC}$ if $f_{\bfC} = 0$. Therefore,
\begin{eqnarray}
L_{\bbU} = L_{\bbC} = \eta L_{\bbC^{ES}},
\end{eqnarray}
for Pauli error models, where $\eta \equiv \abs{\bbC^{ES}}/\abs{\bbC}$ is the proportion of error-sensitive circuits in all Clifford circuits.

The distribution of error-sensitive circuits from Algorithm~\ref{alg:nonU} is non-uniform. Because we uniformly choose slot gates in  $\bar{R}$, the probability of an error-sensitive circuit $\bfC$ is
\begin{eqnarray}
P_{nu}(\bfC) &=& 24^{-(N_R-n)} 8^{-w(\bfC)}24^{-[n-w(\bfC)]} \notag \\
&=& 24^{-N_R} 3^{w(\bfC)}.
\end{eqnarray}
Therefore, the probability of $\bfC$ is proportional to $3^{w(\bfC)}$. If we use Algorithm~\ref{alg:nonU} to sample circuits, we can evaluate $L_{\bbC^{ES}}$ according to
\begin{eqnarray}
L_{\bbC^{ES}} = \eta^{-1} \mathrm{E}[3^{-w(\bfC)} (y_{\bfC} - f_{\bfC})^2]_{nu},
\end{eqnarray}
where the expected value is taken over the distribution $P_{nu}(\bfC)$.

We can generate a uniform distribution of error-sensitive circuits as shown in Algorithm~\ref{alg:U}. In the uniform distribution, the probability of an error-sensitive circuit is $P_{u}(\bfC) = \abs{\bbC^{ES}}^{-1}$. Then, we can evaluate $L_{\bbC^{ES}}$ with $L_{\bbC^{ES}} = \mathrm{E}[(y_{\bfC} - f_{\bfC})^2]_{u}$, where the expected value is taken over the distribution $P_{u}(\bfC)$. By changing the formula of the acceptance probability, we can use the same algorithm to generate other distributions of error-sensitive circuits.

We now summarise the algorithms and analyse their classical-computing costs. Algorithm~\ref{alg:generation} is used to generate an error-sensitive circuit. Provided with an observable $Q$ and a frame with $n$ qubits and $N$ two-qubit gates, Algorithm~\ref{alg:generation} includes operations that conjugate $Q$ (line 3) via $O(N)$ Clifford gates and a conditioned random selection for the single-qubit gates in the first layer (line 5 to 8). The time cost of the conjugating operations is $O(nN)$ according to the efficient simulation algorithm for Clifford gates~\cite{aaronson_improved_2004}, and the time cost of selecting gates in the first layer is $O(n)$. Thus, the cost of Algorithm 1 is $O(nN)$. Algorithm~\ref{alg:nonU} and Algorithm~\ref{alg:U} are used to sample error-sensitive circuits according to the non-uniform distribution $P_{nu}(\bfC)$ and uniform distribution $P_u(\bfC)$, respectively. To generate $N_T$ circuits, the costs for both algorithms are $O(N_T nN)$, because the elementary building block of both algorithms is nothing but the circuit generation given in Algorithm~\ref{alg:generation}, which is repeated for $N_T$ times. The numerical result in Appendix~\ref{sec:veri_feasi} demonstrates that the number of error-sensitive circuits $N_T$ required to perform learning-based error mitigation does not increase (as far as we have observed) with either the number of gates or the number of qubits. Overall, the cost scales linearly with the number of qubits and the number of gates. Noting that the sampling algorithms assume that two-qubit gates are Clifford and errors are independent of single-qubit gates. We give discussion in Appendix~\ref{sec:conditions} about the implementation of the algorithms when the assumptions are not satisfied.

\subsection{Phenomenological error model}
\label{sec:PEM}

In this section, we introduce the phenomenological error model which quantifies the bias caused by realistic errors in a circuit. Then, we show that the phenomenological error model can be effectively represented by a global depolarising model with fluctuation, and the fluctuation is $O(1/\sqrt{N})$ times smaller than the depolarising rate. This result suggests that, if we are able to use error mitigation to cancel the impact of the effective global depolarising error, we can reduce the bias caused by realistic errors by a factor of $O(1/\sqrt{N})$.

Before introducing our phenomenological error model, we give a brief overview of realistic error models. Consider a quantum gate with the unitary operator $U_i$, the error-free output state of the gate is $[U_i]\rho_i$, where $\rho_i$ is the input state. When the gate is imperfect, we can always express the output state with error as $\mathcal{N}_i [U_i]\rho_i$ (assuming the noisy circuit is a Markov process), where the completely positive map $\mathcal{N}_i$ describes the effect of noise associated with the gate. In the global depolarising model, $\mathcal{N}_i = (1-\epsilon)[\openone] + \epsilon\mathcal{D}$. In realistic error models, $\mathcal{N}_i$ is usually caused by local processes, such as dephasing, dissipation and imperfections in the coherent evolution. If the gate acts on qubit-1 and qubit-2, the noise mainly affects these two qubits. Taking a Pauli error model as an example, the noise map reads
\begin{eqnarray}
\mathcal{N}_i = (1-\frac{16\epsilon}{15})[I^{\otimes n}] + \frac{16\epsilon}{15}\mathcal{D}_{1,2},
\label{eq:GateDepol}
\end{eqnarray}
where
\begin{eqnarray}
\mathcal{D}_{1,2} \equiv \frac{1}{16}\sum_{P_1,P_2=I,X,Y,Z}[P_1\otimes P_2\otimes I^{\otimes(n-2)}].
\end{eqnarray}
We call this particular Pauli error model the gate depolarising model, in which probabilities of Pauli errors are the same. We can rewrite this summation-form error model into the product form
\begin{eqnarray}
\mathcal{N}_i = \prod_{P_1,P_2=I,X,Y,Z}\left[(1-p)[I^{\otimes n}]+p[P_1\otimes P_2\otimes I^{\otimes(n-2)}]\right], \notag \\
\quad
\end{eqnarray}
where $p\simeq \epsilon/15$. In the product form, the noise map is a product of $15$ independent maps, and we call each of them a Pauli error channel.

The global depolarising model with fluctuation can characterise the impact of realistic errors in large circuits. Given a circuit $\bfC$, the error-free final state is $\rho_0 = U\ketbra{0}{0}^{\otimes n}U^\dag$. In our error model, the erroneous final state is $\rho = (1-\epsilon_{\bfC})\rho_0 + \epsilon_{\bfC}\rho_m$, where $\epsilon_{\bfC}$ is the circuit-dependent depolarising rate. According to this model, we have $y_{\bfC} = (1-\epsilon_{\bfC})f_{\bfC}$. If we allow $\epsilon_{\bfC}$ to be any value (rather than limited in the interval $[0,1]$), this error model is a general phenomenological error model. Given any $f_{\bfC}$ and $y_{\bfC}$, the corresponding depolarising rate is $\epsilon_{\bfC} = 1 - y_{\bfC}/f_{\bfC}$. Note that the bias is $\epsilon_{\bfC}f_{\bfC}$, which is always finite even when $f_{\bfC} = 0$ and $\epsilon_{\bfC}$ is infinite.

We write the circuit-dependent depolarising rate as two terms, the average and fluctuation, i.e.~$\epsilon_{\bfC} = \epsilon_0 + \delta \epsilon_{\bfC}$, where
\begin{equation}
    \epsilon_0 \equiv \frac{\mean{\epsilon_{\bfC} f_{\bfC}^2}_{\bbU}}{\mean{f_{\bfC}^2}_{\bbU}}
\end{equation}
is the average depolarising rate with the weight $f_{\bfC}^2$, and $\delta \epsilon_{\bfC}$ is the circuit-dependent fluctuation. We characterise the fluctuation with the weighted standard deviation
\begin{equation}
    \Delta \equiv \sqrt{\frac{\mean{\delta \epsilon_{\bfC}^2 f_{\bfC}^2}_{\bbU}}{\mean{f_{\bfC}^2}_{\bbU}}}.
\end{equation}
The key result is that $\Delta$ increases with the gate number as $O(N^{\gamma})$, and $\gamma \approx 0.5$, see Fig.~\ref{fig:model}.

In the rest part of this section, we show theoretically that the standard deviation $\Delta$ is proportional to $\sqrt{N}$ using a Pauli error model. In the next two sections, we introduce an error mitigation protocol inspired by the phenomenological error model , then we verify the scaling behaviour in numerical simulations of the gate depolarising model, composite error models involving Pauli, amplitude damping and coherent errors, and a model with single-qubit-gate dependent errors. The $\sqrt{N}$ scaling is observed in all the error models.

We focus on Pauli errors to analyse the fluctuation in the phenomenological error model. For general errors, we can use Pauli twirling to convert them into Pauli errors. If error mitigation is concatenated with error correction, logical errors after correction are mainly Pauli errors~\cite{bravyi_correcting_2018}. Suppose errors are independent of single-qubit gates, we have the following relations,
\begin{eqnarray}
\mean{f_{\bfC}^2}_{\bbU} &=& \mean{f_{\bfC}^2}_{\bbC} = \eta \mean{f_{\bfC}^2}_{\bbC^{ES}}, \\
\mean{f_{\bfC}y_{\bfC}}_{\bbU} &=& \mean{f_{\bfC}y_{\bfC}}_{\bbC} = \eta \mean{f_{\bfC}y_{\bfC}}_{\bbC^{ES}}, \\
\mean{y_{\bfC}^2}_{\bbU} &=& \mean{y_{\bfC}^2}_{\bbC} = \eta \mean{y_{\bfC}^2}_{\bbC^{ES}},
\end{eqnarray}
where $\bbU$, $\bbC$ and $\bbC^{ES}$ are circuit sets with the same frame. In the above equations, the first equal sign follows because the Clifford group is a unitary-2 design~\cite{dankert_exact_2009,wang_scalable_2021}, and therefore $\mean{\bullet}_{\bbU} = \mean{\bullet}_{\bbC}$ holds if $\bullet$ is a polynomial of degree two in the gate unitaries. The second equal sign is a consequence of  $f_{\bfC}=0$ when $\bfC \notin {\bbC}^{ES}$ and $\eta = \abs{\bbC^{ES}}/\abs{\bbC}$. Using $f_{\bfC} = \pm 1$ for error-sensitive circuits, we can obtain
\begin{eqnarray}
\eta &=& \mean{f_{\bfC}^2}_{\bbU}, \\
\epsilon_0 &=& \mean{\epsilon_{\bfC}}_{\bbC^{ES}}, \\
\Delta &=& \sqrt{\mean{\delta \epsilon_{\bfC}^2}_{\bbC^{ES}}}.
\end{eqnarray}
These relations allow us to study $\epsilon_0$ and $\Delta$ with error-sensitive circuits.

For simplicity, we consider an error model where two-qubit gates are the dominant sources of errors in actual quantum computing devices. We assume that the initialisation, single-qubit gates and measurement are perfect. In a two-qubit gate, we assume that the probability of Pauli errors are the same, i.e.~the gate depolarising model. We use $N'$ to denote the number of two-qubit gates.

The effect of local Pauli errors is equivalent to that of global depolarising errors in error-sensitive circuits. The unitary transformation of a circuit with $N$ gates is $U = U_N\cdots U_1$. If a Pauli error $\sigma$ occurs after the $i$-th gate, the transformation becomes $U' = U_N\cdots U_{i+1} \sigma U_i\cdots U_1 = \sigma'_{\bfC} U$, where $\sigma'_{\bfC} = U_N\cdots U_{i+1} \sigma U_{i+1}^\dag\cdots U_N^\dag$ is the Pauli error propagated to the end of the circuit. Because gates are Clifford, $\sigma'_{\bfC}$ is also a Pauli operator, i.e.~any Pauli error in the circuit is equivalent to a Pauli error at the end of the circuit. If the probability of the Pauli error is $p$, i.e. the error channel is $(1-p)[\openone]+p[\sigma]$, the final state of the circuit is transformed from $\rho_0$ to $(1-p)\rho_0 + p[\sigma'_{\bfC}]\rho_0$. Then there are two cases: If $\sigma'_{\bfC}$ and the Pauli observable $Q$ are commutative, the expected value is preserved under the Pauli error; otherwise, the expected value is changed from $f_{\bfC}$ to $(1-2p)f_{\bfC}$, i.e.~the equivalent depolarising rate is $2p$.

The overall depolarising rate depends on the number of Pauli error channels. Each two-qubit gate contributes $15$ Pauli error channels according to the product form of the Pauli error model. For a circuit with $N'$ two-qubit gates, there are $M = 15N'$ error channels. Let $(1-p)[\openone]+p[\sigma_k]$ be the $k$-th error channel, $(1-p)[\openone]+p[\sigma^\prime_{k,\bfC}]$ is the corresponding error channel at the end of the circuit. We use the binary number $t_k(\bfC)$ to denote whether the $k$-th error channel affect the observable, i.e.~$t_k(\bfC) = 0$ if $\sigma'_{k,\bfC}$ and $Q$ are commutative, and $t_k(\bfC) = 1$ otherwise. Then, the expected value is changed to $\prod_{k=1}^M (1-2p)^{t_k(\bfC)} f_{\bfC}$. The equivalent depolarising rate is
\begin{eqnarray}
\label{eq:equivDepol}
\epsilon_{\bfC} &=& 1 - \prod_{k=1}^M (1-2p)^{t_k(\bfC)} = \sum_{k=1}^M 2t_k(\bfC) p + O(p^2).~~~~~
\label{eq:epsC}
\end{eqnarray}

The average depolarising rate is proportional to the gate number, and the standard deviation is proportional to the square root of the gate number. We can understand this phenomenon as follows. If we choose the circuit randomly from the circuit set, each error channel is switched on and off randomly, i.e.~each $t_k$ takes a random value. Under the assumption that $t_k$ are independent and identically distributed random variables, the distribution of $\epsilon_{\bfC}$ is binomial. Let $P$ be the probability of $t_k = 1$ and neglect $O(p^2)$ terms, the average depolarising rate is $\epsilon_0 \simeq 2pMP$, and the standard deviation is $\Delta  \simeq \sqrt{2pMP(1-P)}$. Note that $M$ is proportional to the gate number.

In large circuits, the global depolarising model with the depolarising rate $\epsilon_0$ is an approximate phenomenological error model. When we sample circuits composed of noisy gates, the circuit plays the role of a sampler, i.e.~the impact of each gate error is a random variable dependent on the circuit configuration. In a certain regime, the total impact is the summation of individual gate errors. When the gate number is larger, the number of random variables in the summation is larger. According to the law of large numbers, the relative standard deviation of the summation decreases with the number of random variables, i.e.
\begin{equation}
    \frac{\Delta}{\epsilon_0} \propto \frac{1}{\sqrt{M}}, \label{eq:sqrtScaling}
\end{equation}
where $M\propto N' \sim N$. Therefore, $\epsilon_{\bfC}$ is in the vicinity of $\epsilon_0$ with a high probability in large circuits.

The analysis above has shown that local gate errors can be represented by a fluctuating global depolarising error, and the ratio of the fluctuation $\Delta$ to the depolarising rate $\epsilon_0$ is in proportion to $1/\sqrt{N}$. This result will be verified by the numerical simulations in the next two sections. We will show that, if the effective global depolarising error is removed by error mitigation, the remaining error (caused by the fluctuation) scales with the gate number as $1/\sqrt{N}$. In addition, we numerically illustrate the error propagation model used in the above analysis. We show that the overall effect of propagated gate errors will become close to the global depolarising error and the relative difference between them decreases as $1/\sqrt{N}$. We leave the numerical result of error propagation to Appendix~\ref{sec:errPro}.

The analysis in this section assumes a small total error rate $pM$. Under this assumption, we can neglect contributions from the second order in Eq.~(\ref{eq:epsC}). In the section of ``Numerical results of the scaling behaviour'', we randomly take total error rates from about $0.003$ to $0.3$, and we observe the $\sqrt{N}$ scaling behaviour. We remark that a modest total error rate is a general requirement of quantum error mitigation~\cite{cai_quantum_2022, qin_overview_2022}. Unlike quantum error correction, which actively detects and corrects errors in the circuit, most quantum error mitigation protocols correct the result by post-processing the noisy experimental data. When the total error rate is high, i.e.~the fidelity approaches zero, the raw data lose the information about the correct quantum state, from which post-processing cannot recover the information. For example, in probabilistic error cancellation, the sampling overhead is exponential in the number of gates given a constant error rate per gate~\cite{temme_error_2017, endo_practical_2018}.

\subsection{Error mitigation according to the phenomenological error model}
\label{sec:PEMI}

\begin{figure}[tbp]
\begin{center}
\includegraphics[width=1\linewidth]{\figpath/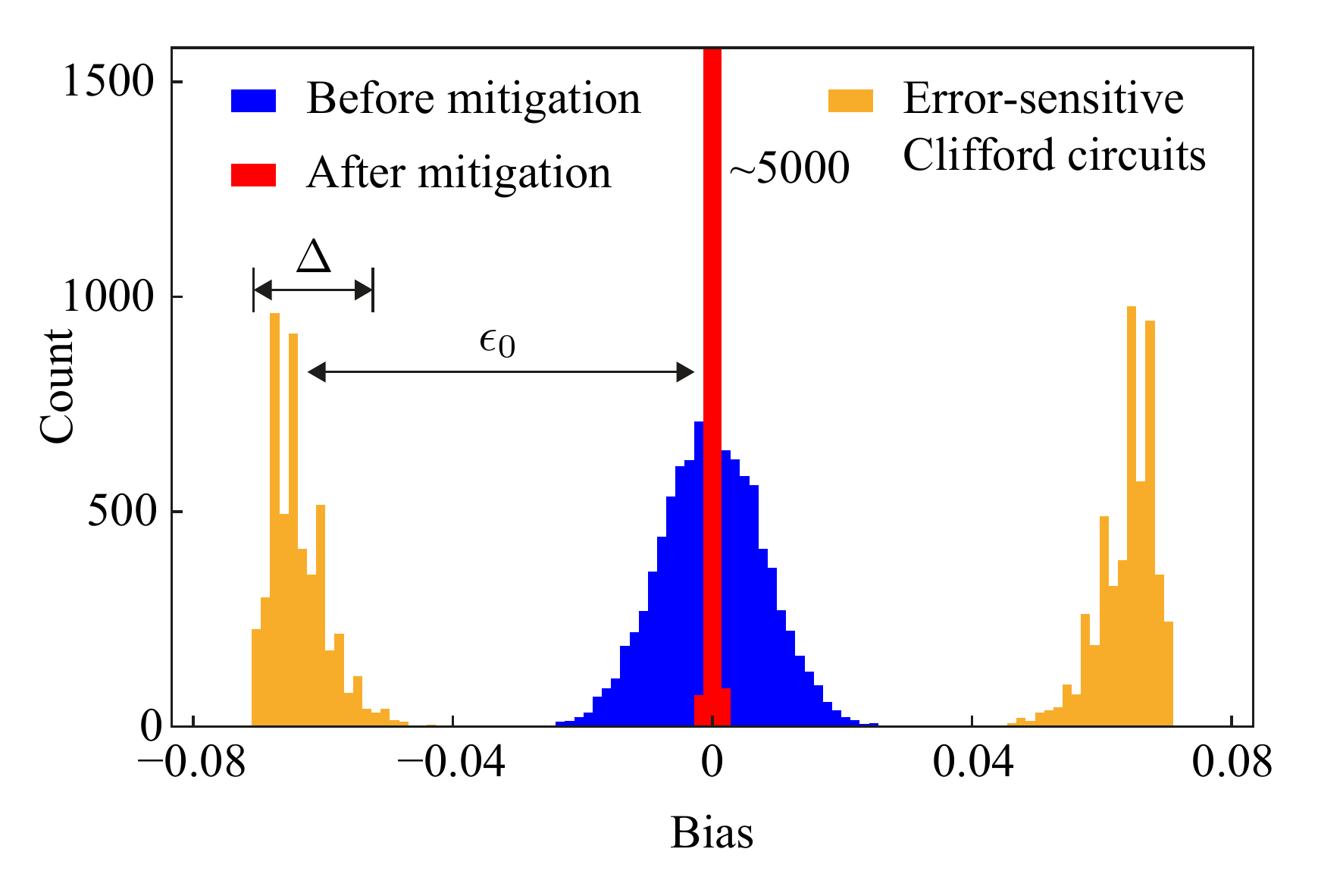}
\caption{
Distributions of the bias for six-qubit periodic-cycling circuits with $72$ two-qubit gates under the gate depolarising noise. The error rate per gate is $0.001$. Before error mitigation, the bias distribution of unitary circuits (the blue histogram) has a shape similar to the Gaussian distribution, and the bias distribution of error-sensitive circuits (the orange histogram) is concentrated at two values. When we mitigate errors according to the average depolarising rate $\epsilon_0$, we move the two peaks to the centre, and the residual bias is determined by the width of the two peaks. Because of the equivalence between the importance Clifford sampling and unitary sampling, the bias of unitary circuits is significantly reduced after error mitigation (the red histogram).
}
\label{fig:distribution}
\end{center}
\end{figure}

According to the phenomenological error model, the effective depolarising rate in large circuits is $\epsilon_0$ with a small fluctuation. We can mitigate errors by compensating the effect of $\epsilon_0$. We use the root mean square error (RMSE) as the measure of the overall accuracy of an error mitigation formula in a circuit set. Before error mitigation, RMSE of unitary circuits with the same frame is $\sqrt{\mean{(y_{\bfC}-f_{\bfC})^2}_{\bbU}} = \sqrt{\eta(\epsilon_0^2 + \Delta^2)} \simeq \sqrt{\eta}\epsilon_0$, which increases linearly with the gate number. Using the error mitigation formula $y'_{\bfC} = (1-\epsilon_0)^{-1}y_{\bfC}$, we can reduce RMSE to $\sqrt{\mean{(y'_{\bfC}-f_{\bfC})^2}_{\bbU}} = (1-\epsilon_0)^{-1} \sqrt{\eta}\Delta \simeq \sqrt{\eta}\Delta$, which increases sublinearly with the gate number. Because $\epsilon_0 = 1 - \mean{y_{\bfC}f_{\bfC}}_{\bbC^{ES}}$, we can measure $\epsilon_0$ (and $\Delta$) by uniformly sampling error-sensitive circuits. Actually, because the fluctuation is small, we can even take $\hat{\epsilon}_0 = 1 - y_{\bfC}f_{\bfC}$ for one randomly generated error-sensitive circuit $\bfC\in \bbC^{ES}$, and it is likely that the error mitigation formula still works. This phenomenological-error-model inspired (PEMI) error mitigation protocol is illustrated in Fig.~\ref{fig:distribution}.

Similar protocols that mitigate errors according to the global depolarising model have been proposed in Refs.~\cite{vovrosh_efficient_2021,urbanek_mitigating_2021,czarnik_error_2020}. In these protocols, the effective depolarising rate is measured in different ways. Before considering general error mitigation formulas, we take the PEMI protocol as an example to verify the phenomenological error model, because the bias of this protocol is directly related to the fluctuation.

In the PEMI protocol, we can further reduce RMSE by optimising the error mitigation formula. If we take
\begin{eqnarray}
y'_{\bfC} = \frac{1-\epsilon_0}{(1-\epsilon_0)^2+\Delta^2}y_{\bfC},
\label{eq:EMFa}
\end{eqnarray}
RMSE after mitigation is reduced to
\begin{eqnarray}
\sqrt{\mean{(y'_{\bfC}-f_{\bfC})^2}_{\bbU}} = \frac{\sqrt{\eta}\Delta}{\sqrt{(1-\epsilon_0)^2+\Delta^2}}.
\end{eqnarray}

\subsection{Numerical results of the scaling behaviour}
\label{sec:numerics}

\begin{figure*}[tbp]
\begin{center}
\includegraphics[width=1\linewidth]{\figpath/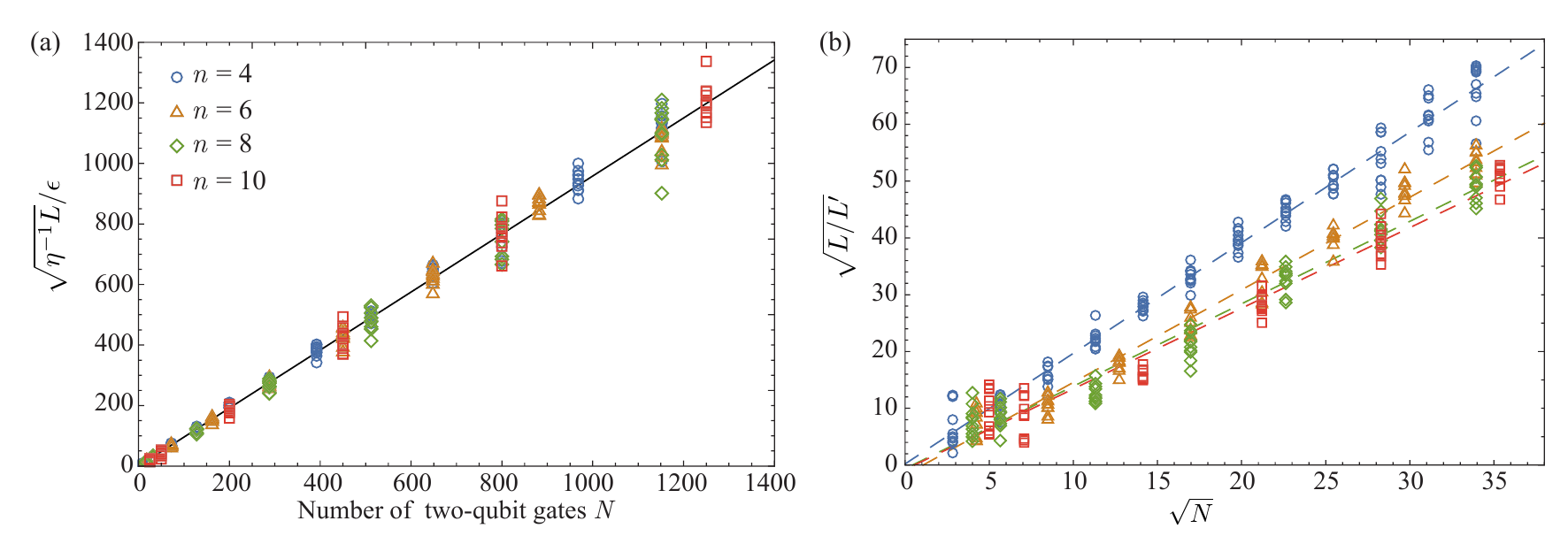}
\caption{
Root mean square errors of linear-network circuits with the gate depolarising model. (a) Root mean square error $\sqrt{L}$ before error mitigation. (b) Root mean square error $\sqrt{L'}$ after error mitigation. In the numerical simulation, we randomly generate a circuit frame with $n$ qubits and $N$ two-qubit gates, and we randomly take the error rate per gate $\epsilon$.
We generate 1000 Clifford circuits according to Algorithm~\ref{alg:nonU} to estimate the phenomenological error rate and then generate 1000 random unitary circuits to compute $L$ and $L^\prime$.}
\label{fig:Linear_depo}
\end{center}
\end{figure*}

\begin{figure*}[tbp]
\begin{center}
    \includegraphics[width=1\linewidth]{\figpath/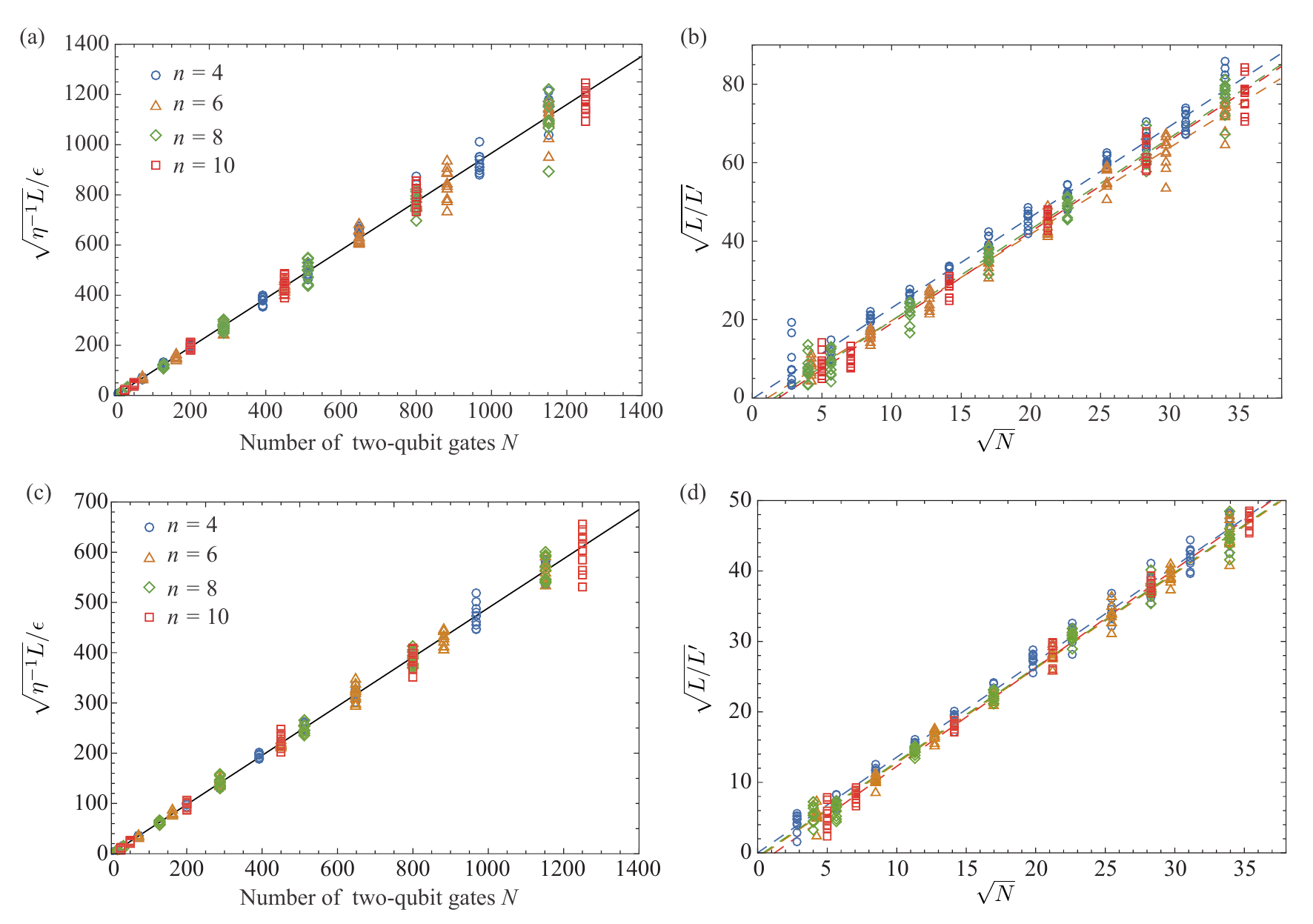}
    \caption{
    Root mean square errors of all-to-all-network circuits. (a) and (c) Root mean square error $\sqrt{L}$ before error mitigation. (b) and (d) Root mean square error $\sqrt{L'}$ after error mitigation. The results in (a) and (b) are obtained with the gate depolarising model, and the results in (c) and (d) are obtained with the composite model. In the numerical simulation, we randomly generate a circuit frame with $n$ qubits and $N$ two-qubit gates, and we randomly take the error rate per gate $\epsilon$. We generate 1000 Clifford circuits according to Algorithm~\ref{alg:nonU} to estimate the phenomenological error rate and then generate 1000 random unitary circuits to compute $L$ and $L^\prime$.
    }
    \label{fig:AlltoAll_comp}
\end{center}
\end{figure*}



\begin{figure}[tbp]
\begin{center}
\includegraphics[width=1\linewidth]{\figpath/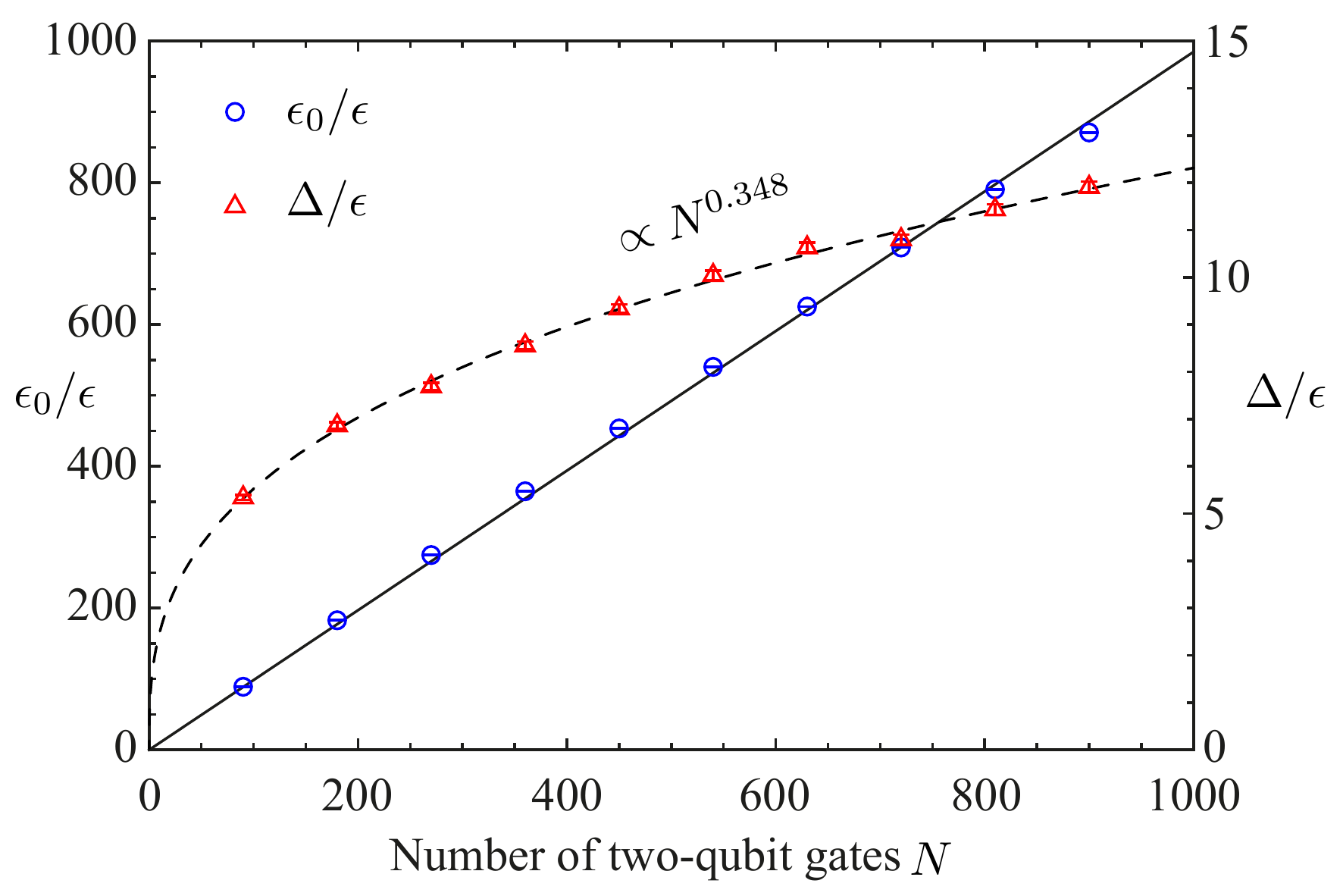}
\caption{
Average depolarising rate $\epsilon_0$ and standard deviation $\Delta$ in six-qubit periodic-cycling circuits. The axis on the left corresponds to $\epsilon_0$ and the axis on the right corresponds to $\Delta$. The error rate per two-qubit gate is $\epsilon=2\times 10^{-4}$, and the error rate of a single-qubit gate $R$ is $0.1\pi^{-1}\epsilon \arccos \frac{\abs{\Tr(R)}}{2}$. The error bar represents one standard deviation.
}
\label{fig:SQGD}
\end{center}
\end{figure}

In this section, we numerically test the PEMI error mitigation formula and verify the scaling behaviour of $\epsilon_0$ and $\Delta$. Results of other error mitigation formulas will be given in the next section.

To demonstrate the scaling behaviour, we generate three families of circuits. In periodic-cycling circuits, two-qubit gates are arranged according to a fixed pattern, and we increase the circuit depth by repeating the pattern. Therefore, periodic-cycling circuits are deterministic. In linear-network circuits, two-qubit gates only act on the nearest neighbouring qubits on a one-dimensional qubit array, and we randomly place two-qubit gates in the circuit. In all-to-all-network circuits, two-qubit gates are also arranged randomly but they can act on any pair of qubits.

We use three types of error models in our numerical calculations: the gate depolarising model with a randomly selected error rate, randomly generated composite error models and a model with single-qubit-gate dependent errors. The gate depolarising model is used to derive the phenomenological error model, but the conclusion holds for other error models. The composite error model involves gate depolarising, dephasing, amplitude damping and coherent errors, which are the typical error sources in actual devices. We generate different composite error models by randomly choosing the weight of each component and observe the same scaling behaviour as the gate depolarising model. The equivalence between Clifford sampling and unitary sampling is also used in deriving the phenomenological error model, which is under the condition that errors are single-qubit-gate independent. In the numerical result, we find that the conclusion on the scaling behaviour holds even if errors are single-qubit-gate dependent. See the Methods section for details of numerical calculations. 

By compensating the average depolarising rate, we can reduce RMSE from $\sqrt{L} = \sqrt{\mean{(y_{\bfC}-f_{\bfC})^2}_{\bbU}} \simeq \sqrt{\eta}\epsilon_0$ to $\sqrt{L'} = \sqrt{\mean{(y'_{\bfC}-f_{\bfC})^2}_{\bbU}} \simeq \sqrt{\eta}\Delta$. According to the discussion in the section of ``Phenomenological error model'', $\epsilon_0 \propto N$ and $\Delta \propto \sqrt{N}$. Therefore, RMSE is reduced in error mitigation by a factor of $\Delta/\epsilon_0 \propto 1/\sqrt{N}$. We verify these scaling behaviours by applying the error mitigation formula in Eq.~(\ref{eq:EMFa}) to randomly generated circuits with up to ten qubits and more than a thousand two-qubit gates. To implement the formula, $\epsilon_0$ and $\Delta$ are measured by sampling error-sensitive circuits. RMSEs before and after error mitigation $\sqrt{L}$ and $\sqrt{L'}$ are calculated and plotted in Figs.~\ref{fig:Linear_depo} and \ref{fig:AlltoAll_comp}. For the model with single-qubit-gate dependent errors, we directly calculate and plot $\epsilon_0$ and $\Delta$ in Fig.~\ref{fig:SQGD}. We can find that numerical results are consistent with scaling behaviours predicted by the phenomenological error model. In addition, we perform experiments on IBM quantum computers~\cite{ibmq} and observe good agreement between the numerical and experimental results. We include the experimental results in Appendix~\ref{app:experiment}.

In Fig.~\ref{fig:AlltoAll_comp}, the error suppression ratio $\sqrt{L/L'}$ for all-to-all-network circuits meets $\sqrt{L/L'}=a\sqrt{N}$ and $a$ is a positive number independent of the qubit number. However, in Fig.~\ref{fig:Linear_depo}, we find that $a$ for linear-network circuits decreases with the qubit number. The difference between all-to-all-network and linear-network circuits is that two-qubit gates in linear-network circuits are short-range, thus it requires more gates for the error on one qubit to propagate across the circuit network.

The error suppression ratio $\sqrt{L/L^\prime}$ are obtained via averaging random unitary circuits, which usually have near-zero expected values. However, in common quantum applications such as variational quantum eigensolver, the expected value is far from zero, which is atypical for random unitary circuits. Thus, we come to ask the question of whether the average suppression ratio of random unitary circuits is also the error suppression ratio of these atypical circuits. To answer this question, we numerically investigate the dependence of the error suppression ratio on the error-free expectation. The numerical result is illustrated in Appendix~\ref{sec:appDepFc}, and the answer is which demonstrates that the average error suppression ratio can be applied to these atypical circuits. 

We note that the $\sqrt{N}$ scaling of error-mitigated result relies on a modest total error rate. This condition is essential for quantum error mitigation methods to work properly~\cite{cai_quantum_2022, qin_overview_2022} and is considered as a general requirement of NISQ computation~\cite{preskill_quantum_2018}. For each data point in Figs.~\ref{fig:Linear_depo} and \ref{fig:AlltoAll_comp}, we randomly choose the error rate per gate $\epsilon$ such that the total error rate $N\epsilon$ is in the interval about $0.003$ to $0.3$. 

\subsection{Error scaling in optimised error mitigation formulas}
\label{sec:ESinQEM}

In this section, we utilise the phenomenological error model to show that one can suppress the scaling of the residual bias in a learning-based manner. For imperfect error extrapolation and probabilistic error cancellation, the error scaling after the optimisation is $\propto \sqrt{N}$. The imperfections are due to the imperfect control of noise in error extrapolation and inaccurate knowledge of the error model in probabilistic error cancellation. For virtual distillation, the result is similar.

First, we analyse the error scaling of error extrapolation. An error mitigation formula usually involves multiple circuits. For each of them, we can effectively characterise the impact of noise using our phenomenological error model. Taking the linear error extrapolation as an example, the two circuits $\bfC_1$ and $\bfC_2$ are the same as the primitive circuit $\bfC$, but the noise level is doubled in $\bfC_2$. In the phenomenological error model of the circuit $\bfC_i$, the average depolarising rate is $\epsilon_i$, the rate fluctuation is $\delta\epsilon_{\bfC,i}$, and the standard deviation is $\Delta_i$. Because $\bfC_1$ and $\bfC_2$ are the same circuit, their fluctuations are correlated: Suppose effective depolarising rates are approximately proportional to the noise level, we have $\epsilon_2 \simeq 2\epsilon_1$ and $\delta\epsilon_{\bfC,2} = 2\delta\epsilon_{\bfC,1}$. Therefore, the fluctuation-caused bias depends on the covariance matrix $K_{i,j} \equiv \eta^{-1}\mean{\delta\epsilon_{\bfC,i}\delta\epsilon_{\bfC,j}f_{\bfC}^2}_{\bbU}$.

\begin{figure*}[tbp]
\begin{center}
\includegraphics[width=1\linewidth]{\figpath/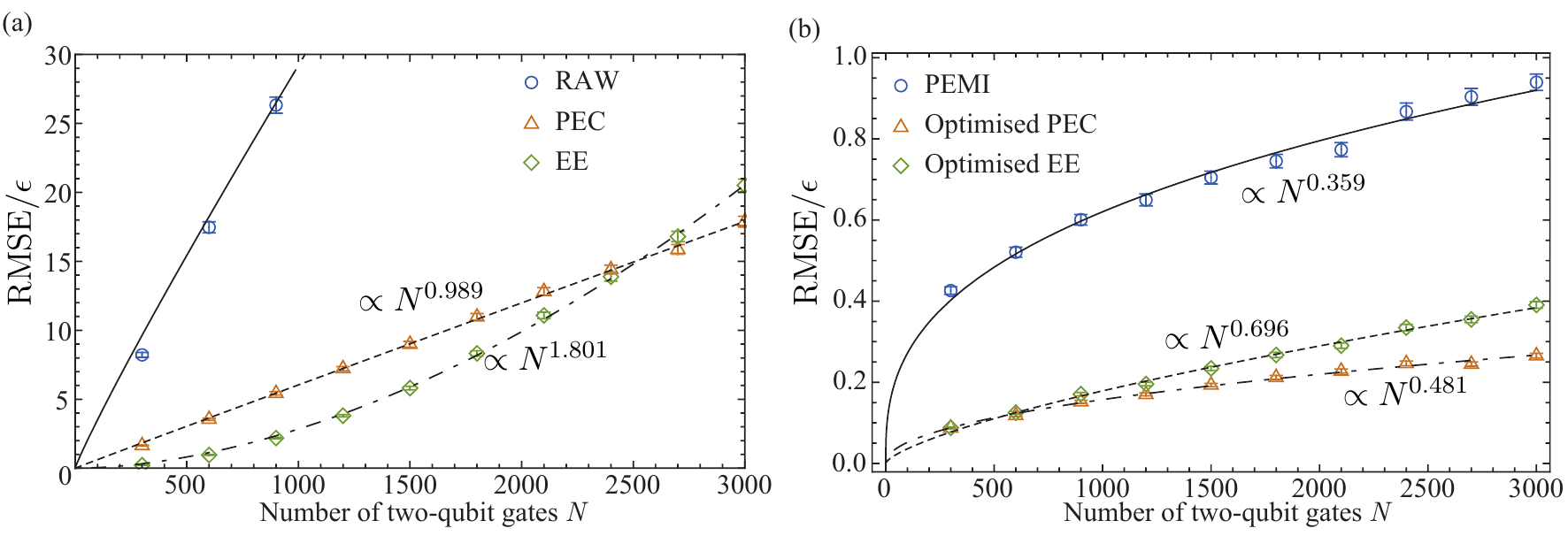}
\caption{
Root mean square errors (RMSE) in error mitigation protocols. The result is obtained using ten-qubit periodic-cycling circuits under the error model in Eq.~(\ref{eq:ErrorModel}) with $\epsilon_d = 8\times 10^{-5}$ and $\epsilon_z = 2\times 10^{-5}$, and we use 1000 Clifford circuits generated via Algorithm~\ref{alg:nonU} for the training and 1000 unitary circuits to compute the  RMSE. The error bar represents one standard deviation. In the raw result without error mitigation, RMSE increases linearly with the gate number. In error extrapolation (EE), noise is increased imperfectly: $\epsilon_d = 1.8\times 10^{-4}$ and $\epsilon_z = 2\times 10^{-5}$ in the error model with a doubled error rate, i.e.~only the gate depolarising component is increased. In probabilistic error cancellation (PEC), we take the inverse map in Eq.~(\ref{eq:inverse}) according to an inaccurate error model with only gate depolarising errors, i.e.~we take $\lambda = -16\epsilon_d/(15-16\epsilon_d)$ before the optimisation and the optimal value after the optimisation.
}
\label{fig:QEMscaling}
\end{center}
\end{figure*}

For the linear extrapolation formula in Eq.~(\ref{eq:EMFlambda}), RMSE after mitigation depends on average depolarising rates $\epsilon_i$ and the covariance matrix $K$, i.e.
\begin{eqnarray}
\sqrt{\mean{(y'_{\bfC}-f_{\bfC})^2}_{\bbU}} = \sqrt{\eta[(E^\dag\Lambda-1)^2+\Lambda^\dag K\Lambda]},
\label{eq:RMSE1}
\end{eqnarray}
where $E = (1-\epsilon_1, 1-\epsilon_2)^{\rm T}$ and $\Lambda = (\lambda, 1-\lambda)^{\rm T}$. Taking $\lambda = \epsilon_2/(\epsilon_2-\epsilon_1)$, we can remove the contribution of average depolarising rates, and RMSE becomes $\sqrt{\mean{(y'_{\bfC}-f_{\bfC})^2}_{\bbU}} = \sqrt{\eta \Lambda^\dag K\Lambda} \leq \sqrt{\eta(\Delta_1^2+\Delta_2^2)}\norm{\Lambda} \propto \sqrt{N}$. Here, we have used that $K$ is positive semi-definite, $\Delta_1^2$ and $\Delta_2^2$ are diagonal elements of $K$, and $\norm{\Lambda} \simeq \sqrt{5}$ does not change significantly with the gate number. Note that this upper bound holds even if the noise is not increased as designed, and we can further reduce RMSE by optimising the parameter $\lambda$. In Fig.~\ref{fig:QEMscaling}, we plot RMSE before and after error mitigation. In the optimised error mitigation formula, we take $\lambda = \epsilon_2/(\epsilon_2-\epsilon_1)$. The numerical result is consistent with the scaling behaviour predicted by the phenomenological error model.

\begin{theorem}
Consider the general extrapolation formula in Eq.~(\ref{eq:EMFL}), let $\epsilon_i$, $\delta\epsilon_{\bfC,i}$ and $\Delta_i$ be the average depolarising rate, rate fluctuation and standard deviation of the circuit $\bfC_i$, respectively, then
$$\min_{\{q_i\}} \sqrt{\mean{(y'_{\bfC}-f_{\bfC})^2}_{\bbU}} \leq \frac{\sqrt{\eta E^\dag K E}}{\norm{E}^2} \leq \frac{\sqrt{\eta\sum_{i}\Delta_i^2}}{\norm{E}}.$$
where $E = (1-\epsilon_1, 1-\epsilon_2, \ldots)^{\rm T}$, $K_{i,j} = \eta^{-1}\mean{\delta\epsilon_{\bfC,i}\delta\epsilon_{\bfC,j}f_{\bfC}^2}_{\bbU}$ and $\eta = \mean{f_{\bfC}^2}_{\bbU}$.
\end{theorem}

The proof is straightforward. Let $\Lambda = (q_1, q_2, \ldots)^{\rm T}$, the expression of RMSE is the same as Eq.~(\ref{eq:RMSE1}). We can prove the theorem by taking $\Lambda = E/\norm{E}^2$.

Second, we investigate the error scaling of probabilistic error cancellation. In probabilistic error cancellation, we reconstruct the transformation of the ideal circuit as a linear combination of transformations of noisy circuits. A practical way is decomposing each ideal gate in the circuit as a linear combination of noisy gates. In general, we can work out the decomposition as follows. If $U_i$ is the unitary operator of the ideal gate, the completely-positive map of the noisy gate is $\mathcal{N}_i[U]$. We can cancel the noise by applying an inverse noise $\widetilde{\mathcal{N}}_i^{-1} = \sum_k q_{i,k} \mathcal{E}_{i,k}$ after the noisy gate, and the overall effective gate is $\widetilde{\mathcal{N}}_i^{-1}\mathcal{N}_i[U]$. Here, $\mathcal{E}_{i,k}$ are some noisy gates, i.e.~we insert the gate $\mathcal{E}_{i,k}$ after the gate $\mathcal{N}_i[U]$ with the quasi-probability $q_{i,k}$. If $\widetilde{\mathcal{N}}_i^{-1} = \mathcal{N}_i^{-1}$, the error in the gate is completely removed; otherwise, effective noise in the gate is $\widetilde{\mathcal{N}}_i^{-1}\mathcal{N}_i$.

We consider a Pauli error model with gate depolarising errors and dephasing errors as an example. For a two-qubit gate on qubit-1 and qubit-2, the noise map is
\begin{eqnarray}
\mathcal{N}_i &=& (1-\frac{16\epsilon_d}{15}-\epsilon_z)[I^{\otimes n}] \notag \\
&&+ \frac{16\epsilon_d}{15}\mathcal{D}_{1,2} + \frac{\epsilon_z}{2}([Z_1]+[Z_2]),
\label{eq:ErrorModel}
\end{eqnarray}
where $Z_i = [I^{\otimes(i-1)}\otimes Z\otimes I^{\otimes(n-i)}]$. Suppose our knowledge about the noise map is inaccurate and we correct the error according to the gate depolarising model, we have
\begin{eqnarray}
\widetilde{\mathcal{N}}_i^{-1} = (1-\lambda)[I^{\otimes n}] + \lambda\mathcal{D}_{1,2}.
\label{eq:inverse}
\end{eqnarray}
When $\lambda = -16\epsilon_d/(15-16\epsilon_d)$ and $\epsilon_z = 0$, we can correct all errors in the gate; otherwise, the effective gate has a finite error rate.

We can suppress the error scaling in imperfect probabilistic error cancellation by optimisation. For an error mitigation formula worked out according to an inaccurate error model, we can treat it as having a virtual quantum computer, in which the error model is given by $\widetilde{\mathcal{N}}_i^{-1}\mathcal{N}_i$. Then, we can describe the error in this virtual machine using the phenomenological error model and reduce the bias using the PEMI protocol. We can use the formula $y^{\prime\prime}_{\bfC} = (1-\epsilon_0^\prime)y^\prime_{\bfC}$, where $\epsilon_0'$ and $y^\prime_{\bfC}$ are respectively the average depolarising rate and expected value in the virtual machine. Then the residual bias of $y''_{\bfC}$ is determined by the standard deviation $\Delta'$ of the virtual machine. Actually, it is not necessary to modify the formula to suppress the error scaling. For example, we can take $\lambda$ in Eq.~(\ref{eq:inverse}) as a variational parameter and optimise it in ICS. The numerical result in Fig.~\ref{fig:QEMscaling} shows that RMSE of probabilistic error cancellation with the optimised $\lambda$ scales as $\propto \sqrt{N}$.

\begin{figure}[tbp]
\begin{center}
\includegraphics[width=1\linewidth]{\figpath/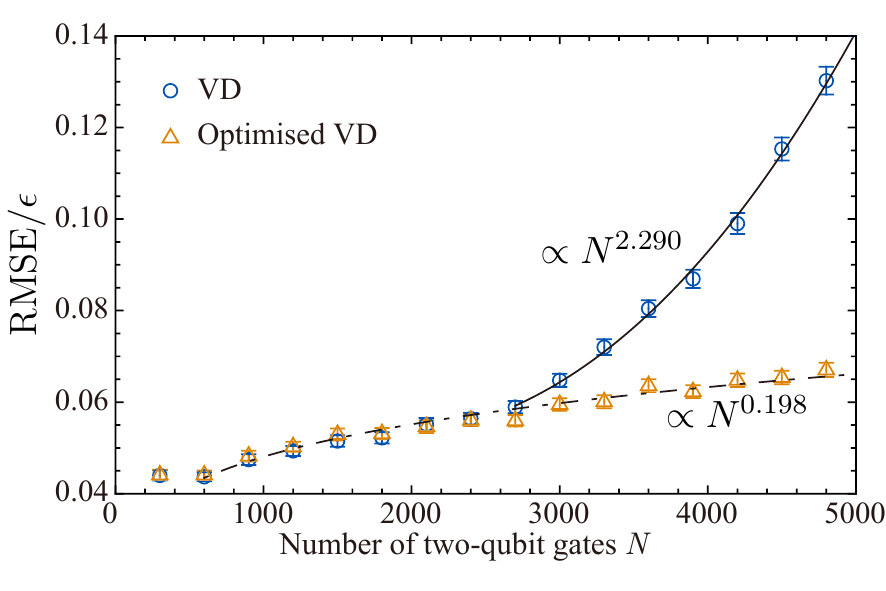}
\caption{
Root mean square errors (RMSE) in virtual distillation (VD) protocols. The error bar
represents one standard deviation. Other details such as the circuit configuration and error model are the same as Fig.~\ref{fig:QEMscaling}.
}
\label{fig:VD}
\end{center}
\end{figure}

Third, we investigate the error scaling of virtual distillation. The virtual distillation formula is nonlinear unlike error extrapolation and cancellation. For a general error mitigation formula, suppose the truncation on the Taylor expansion is valid, we have
\begin{eqnarray}
\label{eq:taylorExpan}
y'_{\bfC} \simeq F(a_1 f_{\bfC_1},a_2 f_{\bfC_2},\cdots) + \sum_i \frac{\partial F}{\partial y_{\bfC,i} } \delta\epsilon_{\bfC,i},
\end{eqnarray}
where $a_i = 1-\epsilon_i$. In Eq.~\eqref{eq:taylorExpan}, we have considered the general error mitigation formula in Eq.~\eqref{eq:generalFormula} and $y_{\bfC,i}=(1-\epsilon_i + \delta \epsilon_{\bfC,i})f_{\bfC_i}$.
If we can remove the zeroth-order term (contribution of average depolarising rates) by taking proper variational parameters in the formula, the bias is determined by fluctuations. For virtual distillation, $F(a_1 f_{\bfC_1},a_2 f_{\bfC_2})=a_1 f_{\bfC_1} / (a_2 f_{\bfC_2})$, therefore, we can compensate average depolarising rates by a factor. In the numerical simulation, we determine the factor by taking the original virtual distillation formula $y'_{\bfC} = y_{\bfC_1}/y_{\bfC_2}$ as a virtual machine and concatenating it with the PEMI protocol according to the formula $y''_{\bfC} = (1-\epsilon_0^\prime)y'_{\bfC}$, where $\epsilon_0^\prime$ is the average depolarising rate of $y^\prime_{\bfC}$. We find that RMSE of the optimised formula scales as $N^\alpha$  and $\alpha < 1/2$ as shown in Fig.~\ref{fig:VD}.

The remaining error after virtual distillation changes from the coherent mismatch~\cite{koczor_exponential_2021} to decoherence error when the gate number increases. With the error-mitigation formula $y^\prime_{\bfC}=\Tr (Q\rho^2)/\Tr (\rho^2)$, the decoherence error is reduced from $N\epsilon$ (gate number times error rate per gate) to $(N\epsilon)^2$, while the coherent mismatch is not suppressed, about which we give a short introduction in Appendix~\ref{sec:appEM}.3. Because the remaining decoherence error increases quadratically with the gate number, the coherent mismatch is the dominant component in the remaining error when the gate number is small, and the decoherence error is the dominant component when the gate number is large. This change in the type of error could explain the bifurcation in Fig.~\ref{fig:VD}, and the result suggests that the optimisation protocol can further reduce the remaining decoherence error but not the coherent mismatch.

In the numerical simulations, we have taken into account imperfect implementations in probabilistic error cancellation and error extrapolation. Assuming the implementation is perfect, probabilistic error cancellation can reduce RMSE to zero, and error extrapolation can reduce RMSE to a much lower level. Note that perfect implementation requires the exact knowledge of the error model or exact control of the error model. In virtual distillation, we have only taken into account errors in those gates that prepare the state $\rho$ and neglected errors in those gates that implement virtual distillation, e.g.~the controlled-swaps in Ref.~\cite{koczor_exponential_2021}.

\section{Discussion}
\label{sec:discussions}

In this work, we show that the residual bias in the computation result after error mitigation scales with the gate number $N$ as $O(\epsilon'N^\gamma)$ if the error mitigation formula is optimised. Here, $\gamma \approx 0.5$, and $\epsilon'$ is a parameter depending on the error rate of quantum gates and the error mitigation formula. In contrast, the bias in the computation result before error mitigation scales linearly with $N$. The two scaling relations lead to a somewhat surprising result: We can suppress the computation error by a larger factor in larger circuits.

In the analysis, we introduce a phenomenological error model characterising errors as the global depolarisation with fluctuation, which captures the impact of realistic noise on the computation result. For the optimisation of an error mitigation formula, we propose ICS as an efficient method of generating training circuits, where only those Clifford circuits sensitive to Pauli errors are selected. The optimised formula removes the average contribution of noise and leaves the fluctuation proportional to $\sqrt{N}$. We verify this result with the numerical simulation of various circuits, error models and error mitigation formulas, from which we observe that the scaling behaviour is universal.

Despite the encouraging scaling of bias in error mitigation, we point out that the circuit size is still limited by the quality of quantum devices. On a quantum device with a finite error rate per gate, the bias increases with the circuit size. Although the bias scaling after error mitigation is advantageous in comparison with the linear error accumulation before mitigation, at certain circuit sizes the computation result becomes sufficiently random that error mitigation cannot faithfully recover the information. Therefore,  the efficacy of error mitigation is conditional on the quality of the quantum device. In general, the minimum requirement for error mitigation to take effect is a non-zero fidelity between the error-free and erroneous circuits, and the performance is better with higher fidelity. Beyond this, the impact of the unmitigated error rate on the accuracy of the mitigated result depends on the mitigation method. In probabilistic error cancellation, for example, the variance in calculating the expectation value of the result increases with the error rate. Another example is that, after the virtual distillation using two copies, the bias in the expectation value scales quadratically with the error rate. Once the device can implement the circuit with sufficiently high fidelity (which is not necessarily close to one but we take a fidelity of $0.9$ as an example), error mitigation can improve the computation result to a much higher accuracy (equivalent to quantum computing with fidelity of $0.99$ if the error is reduced by a factor of ten).

In scalable quantum computers, we can adopt quantum error correction to increase the fidelity of logical qubits. Protocols concatenating error correction with error mitigation have been proposed recently~\cite{suzuki_quantum_2021,piveteau_error_2021,lostaglio_error_2021}. Fault-tolerant devices will enable the implementation of much deeper circuits than NISQ hardware. Our result of the scaling behaviours suggests that error mitigation can perform even better in the fault-tolerant regime than in the NISQ regime.

\section{Methods}

\subsection{Circuits}

We use three families of circuits: periodic-cycling circuits, linear-network circuits and all-to-all-network circuits.

Periodic-cycling circuits. The qubit array has $n$ qubits, and $n$ is even. All qubits are initialised in the state $\ket{0}$. After initialisation, a layer of single-qubit gates is placed, see Fig.~\ref{fig:frames}(a). The circuit pattern is periodic, and each period has two layers of two-qubit gates. In the first layer, a controlled-$Z$ gate is applied on qubit-$(2i-1)$ and qubit-$(2i)$, where $i=1,2,\ldots,n/2$. In the second layer, a controlled-Z gate is applied on qubit-$(2i-1)$ and qubit-$(2i-2)$, and qubit-$0$ and qubit-$n$ are the same qubits. After each two-qubit gate, a single-qubit gate is applied to each of the two qubits. The observable $O$ is $Z$ of the first qubit. All single-qubit gates are taken as slots in the corresponding circuit frame.

Linear-network circuits. Except for the pattern of two-qubit gates and observable, the setup is the same as periodic-cycling circuits. All two-qubit gates are controlled-Z gates. For each of them, we randomly generate an integer $i\in [1,n]$ and apply the two-qubit gate on qubit-$(i-1)$ and qubit-$i$, see Fig.~\ref{fig:frames}(b). The observable is $O = P_1\otimes P_2\otimes\cdots\otimes P_n$, where $P = I,Z$ is taken randomly.

All-to-all-network circuits. It is similar to linear-network circuits. For each of the two-qubit gates, we randomly generate two different integers $i,j\in [1,n]$ and apply the two-qubit gate on qubit-$i$ and qubit-$j$, see Fig.~\ref{fig:frames}(c).

\subsection{Error models}
\label{sec:appErrorModels}
Several error models are used in the numerical simulations.

Gate depolarising model. The model is given in Eq.~(\ref{eq:GateDepol}), and only two-qubit gates have errors. This model is used to generated data shown in Figs.~\ref{fig:model}, \ref{fig:distribution} and \ref{fig:Linear_depo}. In Figs.~\ref{fig:model} and \ref{fig:distribution}, we take $\epsilon = 0.001$. In Figs.~\ref{fig:Linear_depo}, for each data point, we randomly generate a circuit (and the corresponding circuit frame) and an error rate. For a circuit with $N$ two-qubit gates, we generate a random real number $\eta\in[-2.5,-0.5]$, and we take $\epsilon = 10^{\eta}/N$ as the error rate per gate. Notice that $10^{\eta}$ is the total error rate. 

Composite error model. Only two-qubit gates have errors. For a two-qubit gate $U$, the gate with errors is
$$\mathcal{A}_{2}\mathcal{A}_{1}[R_{2,Z}R_{2,Y}R_{2,X}][R_{1,Z}R_{1,Y}R_{1,X}]\mathcal{Z}_2\mathcal{Z}_1\mathcal{N}[U],$$
where $\mathcal{N}$ is the gate depolarising error in Eq.~(\ref{eq:GateDepol}) with the error rate $\epsilon_d$, $\mathcal{Z}_i = (1-\epsilon_{i,z})[I]+\epsilon_{i,z}[Z]$ is the dephasing error on qubit-$i$, $R_{i,P} = e^{-i\frac{\theta_{i,P}}{2}P}$ is a single-qubit rotation on qubit-$i$, and
\begin{eqnarray}
\mathcal{A}_{i} &=& \left[\frac{\openone+Z}{2} + \sqrt{1-\epsilon_{i,a}}\frac{\openone-Z}{2}\right] \notag \\
&&+ \left[\sqrt{\epsilon_{i,a}}\frac{X+iY}{2}\right]
\end{eqnarray}
is the amplitude damping on qubit-$i$. This model is used to generate data shown in Fig.~\ref{fig:AlltoAll_comp} (c) and (d). For each data point, we randomly generate the error model parameters as follows. For a circuit with $N$ two-qubit gates, we generate a random real number $\eta\in[-2.5,-0.5]$, and we take $\epsilon = 10^{\eta}/N$ as the error rate per gate. Then, we take $\epsilon_d = (1+0.2\kappa_d)\epsilon/9$, $\epsilon_{i,z} = (1+0.2\kappa_{i,z})\epsilon/9$, $\theta_{i,P} = \kappa_{i,P}\epsilon/9$ and $\epsilon_{i,a} = (1+0.2\kappa_{i,a})\epsilon/6$. Each $\kappa$ is taken randomly in the interval $[-1,1]$.

Gate-dependent error model. In this model, both single-qubit and two-qubit gates have errors. The error model is the gate depolarising model. For two-qubit gates, the noise map is given by Eq.~(\ref{eq:GateDepol}). For a single-qubit gate $R$, the gate with error is $\mathcal{S}[R]$, where
\begin{eqnarray}
\mathcal{S} = (1-\frac{4\epsilon_s}{3})[I] + \frac{\epsilon_s}{3} \sum_{P=I,X,Y,Z}[P],
\end{eqnarray}
and $\epsilon_s = 0.1\pi^{-1}\epsilon \arccos \frac{\abs{\Tr(R)}}{2}$. This model is used to generate data shown in Fig.~\ref{fig:SQGD}, and we estimate $\epsilon_0$ and $\Delta$ using $10000$ unitary circuits in $\bbU$.

Gate depolarising and dephasing model. The model is given in Eq.~(\ref{eq:ErrorModel}), and only two-qubit gates have errors. This model is used to generate data shown in Figs.~\ref{fig:QEMscaling} and \ref{fig:VD}. In the numerical simulation, we approximate the error model with $\mathcal{Z}_2\mathcal{Z}_1\mathcal{N}$ for simplicity in coding, which only causes a small difference and will not change the conclusion.

The above error models take into consideration kinds of physical noise processes and are able to simulate noises in realistic quantum devices. The depolarising error $\mathcal{N}$ and dephasing error $\mathcal{Z}$ simulates the relaxation process and the dephasing process~\cite{ioffe_asymmetric_2007, wang_single_2021}, which are the main contributions to noise in realistic quantum devices. Amplitude damping $\mathcal{A}$ refers to the infidelity caused by energy dissipation. Random rotations $R$ refer to coherent errors caused by imperfect controls. This composite model takes into consideration all the above realistic imperfections and it was demonstrated in Ref.~\cite{wang_scalable_2021} that the composite model can produce error distributions resembling that in experiments on a superconducting quantum processor. The single-qubit-gate dependent error model $\mathcal{S}$ is the single-qubit depolarising error with an error rate depending on the gate parameters. This error model takes into consideration the realistic situation that gate errors increase with the gate time. Additionally, we will make a direct comparison between the experimental results and simulation results in Appendix~\ref{app:experiment} and show that experimental results are consistent with simulation results.

\subsection{Error mitigation protocols}
\label{sec:appEMprotocols}

We verified the scaling behaviour by simulating various error mitigation protocols. The formula in Eq.~(\ref{eq:EMFa}) is used to generate data shown in Fig.~\ref{fig:distribution}, \ref{fig:Linear_depo} and \ref{fig:AlltoAll_comp}. The PEMI protocol in Fig.~\ref{fig:QEMscaling} is $y'_{\bfC} = (1-\epsilon_0)^{-1}y_{\bfC}$. In optimised error extrapolation, we take $\lambda = \epsilon_2/(\epsilon_2-\epsilon_1)$. In optimised probabilistic error cancellation, we take $\lambda = -16\epsilon_d/(15-16\epsilon_d)-2\epsilon_z$: We have searched for the optimal $\lambda$ using ICS data and found that the numerical optimal value is close to it. In optimised virtual distillation in Fig.~\ref{fig:VD}, the formula is $y''_{\bfC} = (1-\epsilon'_0)^{-1}y'_{\bfC}$. To implement optimised error mitigation formulas, we estimate $\epsilon_0$, $\Delta$, $\epsilon_1$, $\epsilon_2$ or $\epsilon'_0$ using $1000$ error-sensitive circuits, according to Algorithm~\ref{alg:nonU}. Then, we generate $1000$ unitary circuits with the same frame to estimate RMSE.

\section*{Data Availability}
The data that support the findings of this study are available from the corresponding author upon reasonable request.

\section*{Code Availability}
The codes that support the findings of this study are available from the corresponding author upon reasonable request.

\section*{Acknowledgements}
We thank Hang Ren for the discussions. We acknowledge the use of simulation toolkit QuESTlink~\cite{jones_questlink_2020} for this work. We acknowledge the use of IBM Quantum services for this work. DYQ and YL are supported by National Natural Science Foundation of China (Grants No. 11875050 and No. 12088101) and NSAF (Grant No. U1930403). YC acknowledges support from US Department of Energy (Award No. DE-SC0019318).

Note.---When preparing the manuscript, we notice a recent preprint arXiv:2111.14907 that reports the global depolarising model as an effective model of noisy quantum circuits. This work studies the distribution of measurement outcomes in circuits with single-qubit noise channels. In comparison, our work studies expected-value computing using circuits with two-qubit noise channels as the dominant error source. We focus on properties of circuits with the same circuit frame, and we use the effective model in error mitigation. Our final result is on the bias scaling of error mitigation formulas.

\section*{Author Contributions}
DYQ, YC and YL together conceived the ideas. YL developed the theory. DYQ and YL performed the numerical simulation. DYQ and YC implement the experiment. DYQ, YC and YL prepared the manuscript.

\section*{Competing Interests}
The authors declare no competing interests.

\appendix

\section{Error mitigation}
\label{sec:appEM}

Quantum noise is the main obstacle preventing us from implementing desired quantum computations. Error mitigation refers to the recently-proposed techniques to handle noises with low quantum expenses, including error extrapolation, probabilistic error cancellation and virtual distillation etc. We here give a brief introduction to some of those techniques.

\subsection{Error extrapolation}
Suppose the erroneous expected value can be expressed as the series expansion
\begin{eqnarray}
    y_{\bfC_i} = f_{\bfC} + \sum_{k=1}^m a_k (r_i \epsilon)^k + O((r_i \epsilon)^{m+1}),\label{eq:polyFit}
\end{eqnarray}
where the coefficient $a_k$ is independent of $i$ if we assume that the generator of noise operation is independent of the noise amplification factor $r_i$. We can infer the error-free expectation $f_{\bfC}$ by taking Eq.~(\ref{eq:polyFit}) as the polynomial fitting function and extrapolating to the zero error limit $r=0$. Taking Richardson extrapolation~\cite{temme_error_2017} as an example, the error extrapolation formula reads
\begin{eqnarray}
    y_{\bfC}^\prime = \sum_{i=1}^{m+1} q_i y_{\bfC_i},
\end{eqnarray}
where the coefficients are determined by
\begin{equation}
    \sum_{i=1}^{m+1} q_i = 1, \  \sum_{i=1}^{m+1} q_i r_i^k = 0, \  \forall \ k=1,2,\dots,m
\end{equation}
and the remaining error is $\vert y_{\bfC}^\prime - f_{\bfC} \vert = O(\epsilon^{m+1})$.
Take $r_1=1,r_2=2$ and $m=1$, it becomes the linear extrapolation as in Eq.~\eqref{eq:LEF} of the main text.

\subsection{Probabilistic error cancellation}
As we provided in Eq.~\eqref{eq:PEC} of the main text, the error-free map in  probabilistic error cancellation is expressed as
\begin{eqnarray}
[U] = \sum_i q_i \mathcal{E}_i,
\end{eqnarray}
where $q_i$ are quasi-probabilities and $\mathcal{E}_i$ is the map of a noisy circuit $\bfC_i$. If there is only local gate error, e.g. there is no cross-talk between gates, probabilistic error cancellation can be conducted in a gate-wise manner. Suppose the erroneous map of the $j$-th gate is $\mathcal{M}_j = \sum_k p_k [\sigma_k][U_j]$ and $p_k$ is already determined using gate-set tomography, we can realise the error-free gate via
\begin{eqnarray}
    [U_{j}] = \sum_i q_{i,j} \mathcal{E}_{i,j},
\end{eqnarray}
where $\mathcal{E}_{i,j} = [\sigma_i]\mathcal{M}_j$ is realised by inserting Pauli gate $[\sigma_i]$ in front of the $j$-th gate and $q_{i,j}$ is the solution to
\begin{eqnarray}
    \left(\sum_i q_{i,j} [\sigma_i]\right)\left(\sum_k p_k [\sigma_k]\right) = [\openone].
\end{eqnarray}
Then the formula for error mitigated circuit is
\begin{eqnarray}
    [U] &=& \prod_{j=1}^N [U_j] \notag \\
        &=& \sum_i (\prod_{j=1}^n q_{i,j})(\prod_{j=1}^n \mathcal{E}_{i,j}).
\end{eqnarray}

\subsection{Virtual distillation}
\label{sec:appVD}
The objective of virtual distillation is to obtain the expectation of an observable in the purified state $\rho^\prime = \rho^k / \text{Tr}(\rho^k)$. Suppose the spectrum decomposition of the erroneous state is $\rho=\sum_{i=0}^{2^n-1}\lambda_i\ket{\psi_i}\bra{\psi_i}$, the purified state reads
\begin{equation}
    \rho^\prime = \frac{\ket{\psi_0}\bra{\psi_0}+\sum_{i=1}^{2^n-1}(\lambda_i/\lambda_0)^k\ket{\psi_i}\bra{\psi_i}}{1+\sum_{i=1}^{2^n-1}(\lambda_i/\lambda_0)^k}.
\end{equation}
where $\lambda_0$ is the largest eigenvalue. $\rho^\prime$ exponentially gets close to $\ket{\psi_0}\bra{\psi_0}$ as $k$ increases, and error can be completely removed in the limit of $k=\infty$ if $\ket{\psi_0}\bra{\psi_0}$ is the true error-free state $\rho_0$. However, $\ket{\psi_0}\bra{\psi_0}$ could deviate from $\rho_0$ even when the error is completely incoherent, thus virtual distillation usually has an additional error caused by the coherent mismatch $1-\text{Tr}(\rho_0\ketbra{\psi_0}{\psi_0})$~\cite{koczor_exponential_2021}.

\begin{figure}[tbp]
\begin{center}
\includegraphics[width=1\linewidth]{\figpath/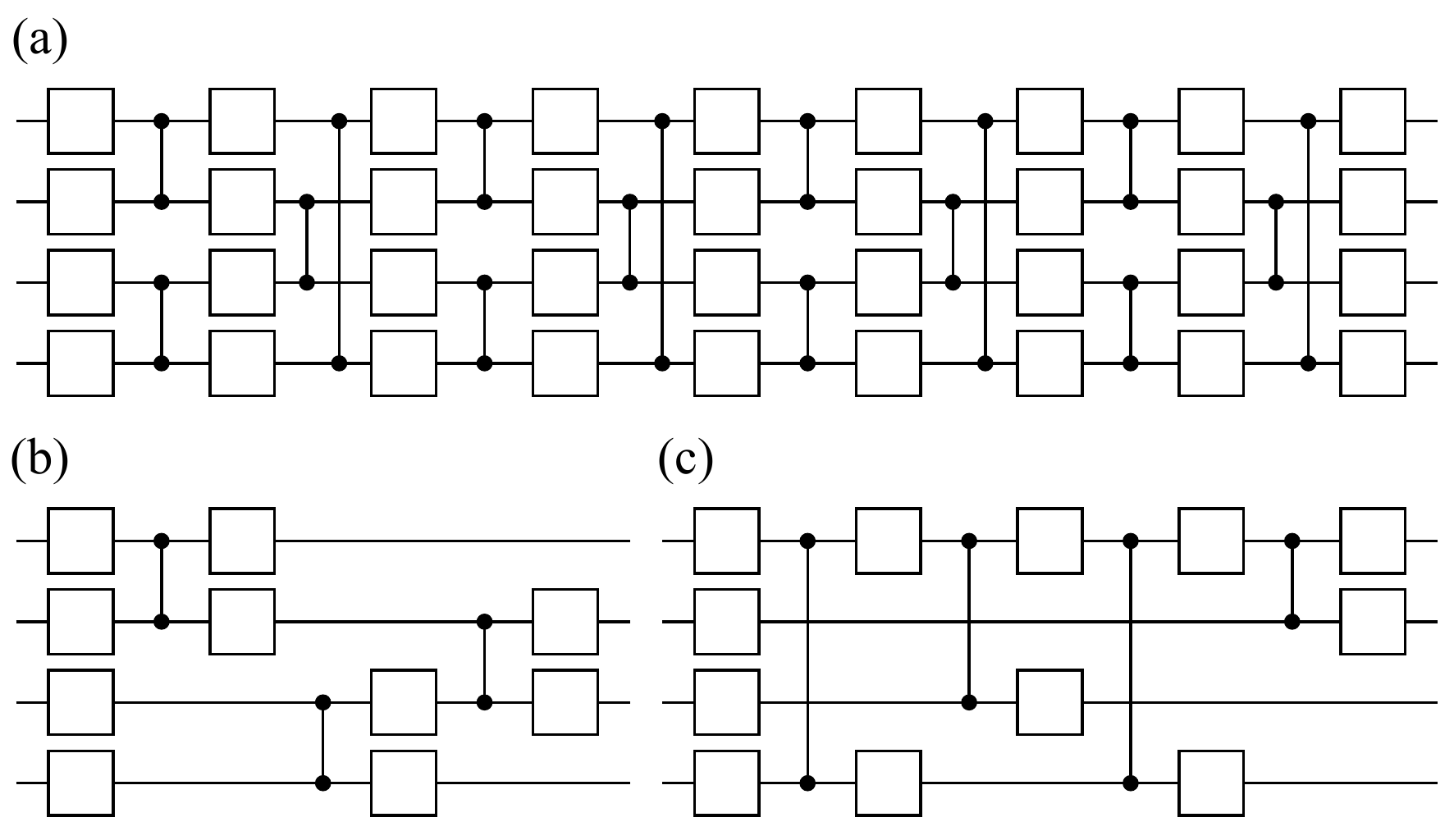}
\caption{
Example of circuit frames. (a) periodic-cycling frame. (b) linear frame. (c) all-to-all frame. Each square represents a single-qubit gate.
}
\label{fig:frames}
\end{center}
\end{figure}

\begin{figure}[tbhp]
\begin{center}
\includegraphics[width=1\linewidth]{\figpath/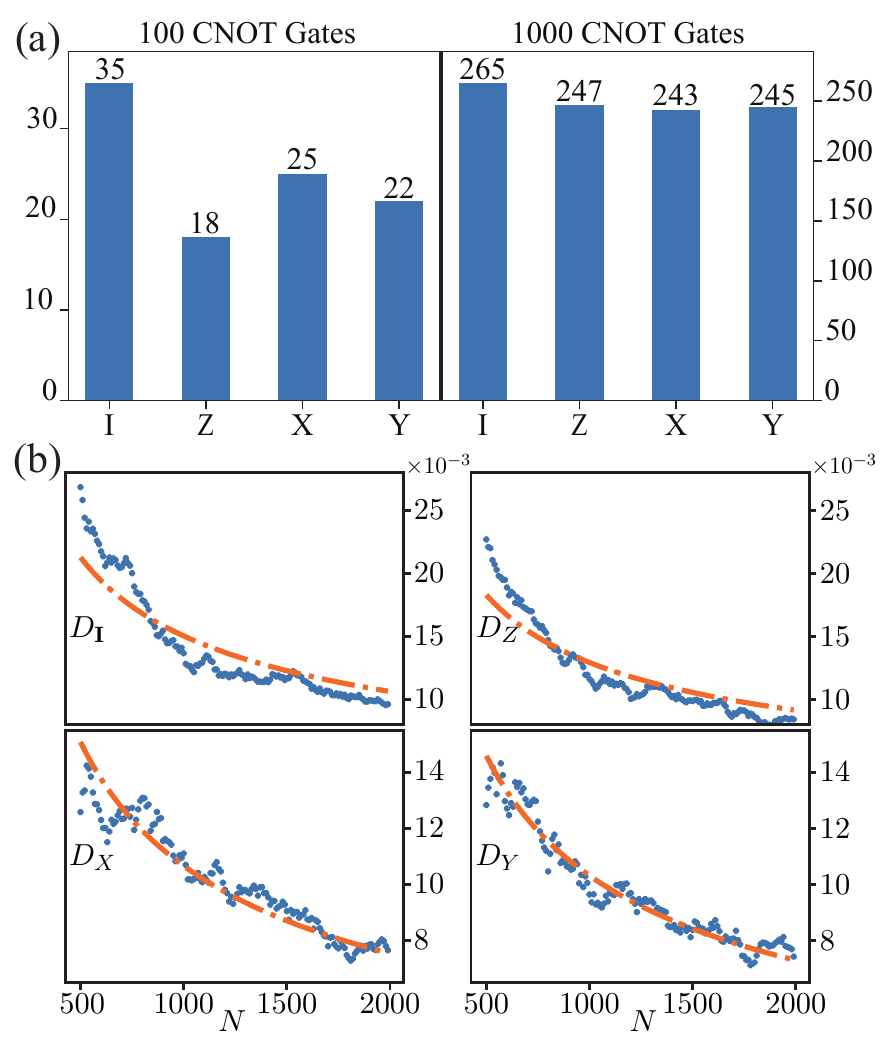}
\caption{
Numerical demonstration of converging to the global depolarising error. We consider a ten-qubit periodic-cycling Clifford circuit consisting of $N$ CNOT gates, and the error on a CNOT gate is $\sigma_i = X\otimes X, \  \forall i=1,2,...,N$, i.e.~a Pauli $X$ error occurred on each qubit of the CNOT gate.  When the number of gates increases, the overall propagated error converges to the global depolarising error. (a) The empirical counts of propagated errors on the first qubit. It visualises that the counts of different Pauli errors $N_\mathbb{I},N_X,N_Y,N_Z$ are relatively closer to $N/4$ when $N$ is larger. (b) $D_{\sigma^\prime}$ is defined as $D_{P^\prime} = \text{E}[\vert \frac{N_{P^\prime}}{N} - \frac{1}{4} \vert]$ and $\text{E}[\cdot]$ denotes expectation over Clifford circuits with the same frame. In the plots, $D_{P^\prime}$ is estimated via averaging over $1000$ Clifford circuits, and it shows that the estimated $D_{P^\prime}$ (blue dots) fit $1/\sqrt{N}$ functions (orange dotted-dash curves).}
\label{fig:GlobalDepol}
\end{center}
\end{figure}

\begin{figure}[tbhp]
\begin{center}
\includegraphics[width=1\linewidth]{\figpath/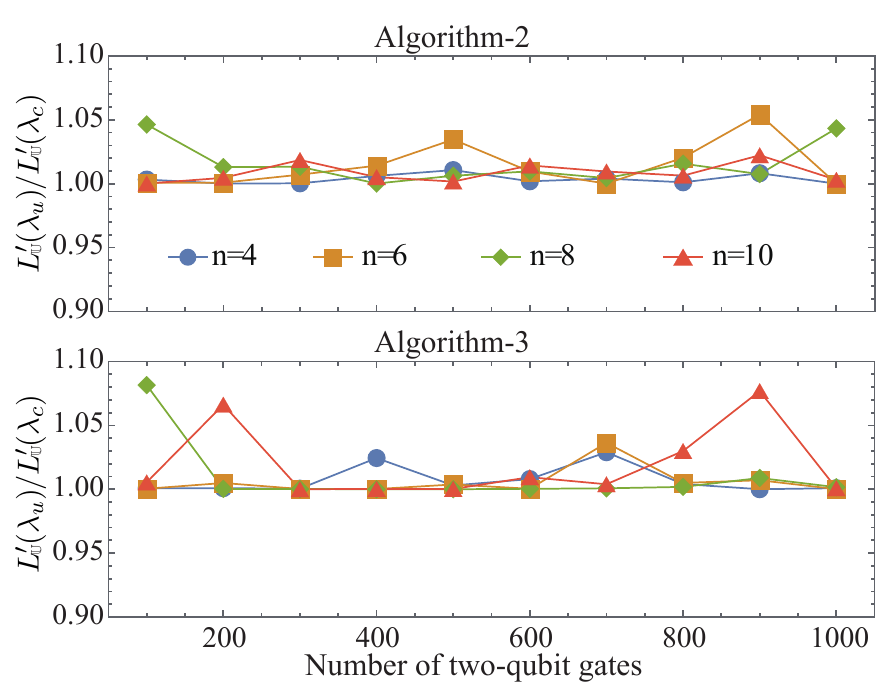}
\caption{
Verification of feasibility of the importance Clifford sampling. The error mitigation formula used here is the PEMI formula in Eq.~(\ref{eq:EMFa}) of the main text. The error model is the gate depolarising model, and the error rate per gate is $2\times 10^{-4}$. $\lambda_u$ and $\lambda_c$ are calculated via sampling $1000$ random unitary and error-sensitive Clifford circuits, respectively. The circuit frame is the all-to-all network, and the number of two-qubit gates is $1000$. Algorithm~\ref{alg:nonU} and Algorithm~\ref{alg:U} are used to sample Clifford circuits for the upper plot and lower plot, respectively. The plots show that $L_\mathbb{U}^\prime(\lambda_c) / L_\mathbb{U}^\prime(\lambda_u)$ does not increase with either the number of gates or qubits.}
\label{fig:costScaling}
\end{center}
\end{figure}

\begin{figure}[tbhp]
\begin{center}
\includegraphics[width=1\linewidth]{\figpath/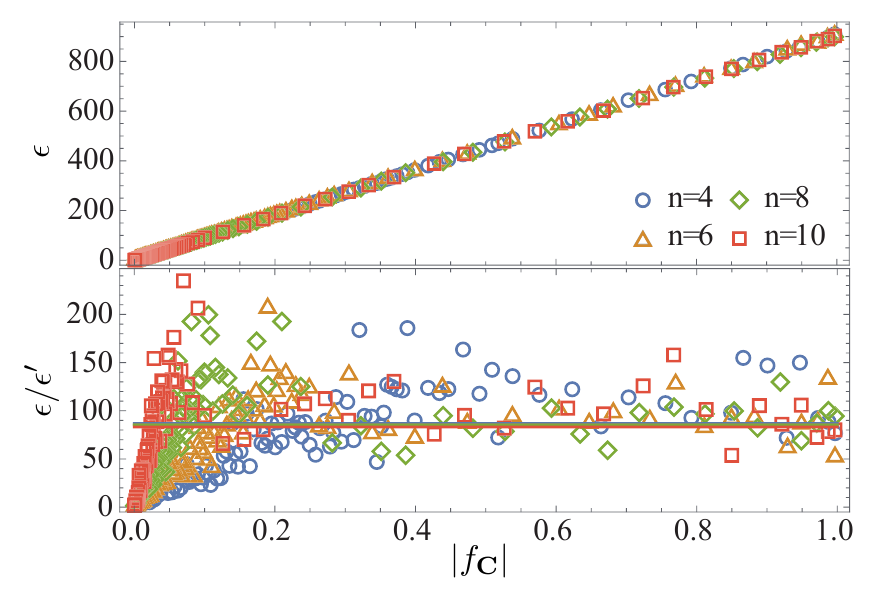}
\caption{
The dependence of errors before and after error mitigation on the error-free computation result. $\vert f_{\bfC} \vert$ is the absolute error-free computation result that ranges from 0 to 1. $\epsilon$ and $\epsilon^{\prime}$ are respectively the errors before and after error mitigation. Here, we take the PEMI formula in error mitigation. We generate 1000 error-sensitive circuits to determine the parameter in the formula and perform error mitigation on 1000 random unitary circuits. Each data point in the plots is obtained via averaging ten random unitary circuits with $\vert f_{\bfC} \vert$ values in the vicinity. The horizontal line in the lower panel represents the value of $\epsilon/\epsilon^\prime$ obtained from random error-sensitive Clifford circuits, whose expectation equals the value of $\epsilon/\epsilon^\prime$ obtained from random unitary circuits. The error model is the gate depolarising model, and the error rate per gate is $2\times 10^{-4}$.}
\label{fig:fc_dependence}
\end{center}
\end{figure}

\section{Error propagation}
\label{sec:errPro}

In the ``Phenomenological error model'' section, the analysis shows that the impact of realistic errors can be described as fluctuating global depolarising error. Suppose $\epsilon_0$ is the effective depolarising rate and $\Delta$ is the fluctuation, we have concluded that $\Delta / \epsilon_0 \sim 1/\sqrt{N}$ where $N$ is the number of gates. Here, we provide an alternative approach to justify the fluctuating global depolarising error model. We consider a Clifford circuit $U^\prime = U_N \cdots U_{i+1}\sigma_i U_i \cdots U_1 = \sigma^\prime_i U$ with a Pauli error $\sigma_i$ occurred at the $i$-th gate, where $U=U_N U_{N-1} \cdots U_1$ is the error-free circuit and $\sigma^\prime_i = U_N \cdots U_{i+1} \sigma_i U^\dagger_{i+1} \cdots U^\dagger_N$ is the propagated error. For Clifford circuits, $\sigma_i^\prime$ can be expressed as $\sigma^\prime_i = \pm P^\prime_{i,1}\otimes P^\prime_{i,2}\otimes\cdots\otimes P^\prime_{i,n}$ and $P_{i,j}^\prime\in\{I,X,Y,Z\}$ for all $j=1,2,...,n$. From now on, we focus on the measurement on the first qubit. Suppose the probability of $\sigma_i$ is $p$. The full expression of the propagated error on the first qubit is $(1-p)[\mathbb{I}]+p[P^\prime_{i,1}]$. If the Pauli error may occur in every gate, the overall propagated error on the first qubit is
\begin{eqnarray}
    \mathcal{G} &=& \prod_{i=1}^N \left((1-p)[I]+p[P_{i,1}^\prime]\right) \notag\\
     &=& \prod_{P^\prime \in \{\mathbb{I},X,Y,Z\}} \left((1-p)[I]+p[P^\prime]]\right)^{N_{P^\prime}},
\end{eqnarray}
where $N_{P^\prime}$ is the number of errors propagated to $\pm P^\prime\otimes\cdots$. If $N_X = N_Y = N_Z$, the overall propagated error is exactly the global depolarising error. We now numerically demonstrate that $\frac{N_{P^\prime}}{N}\approx\frac{1}{4}\quad \forall\quad P^\prime \in \{I,X,Y,Z\}$ when the gate number is large, and the expectation of $\vert \frac{N_{P^\prime}}{N} - \frac{1}{4}\vert$ decrease as $\frac{1}{\sqrt{N}}$. The result is shown in Fig.~\ref{fig:GlobalDepol}, which verifies our result in Eq.~(\ref{eq:sqrtScaling}) of the main text.

\section{Verification of feasibility}
\label{sec:veri_feasi}

We have proposed optimising the error mitigation formula via ICS. Here, we verify that the optimisation via ICS is feasible with increasing numbers of gates or qubits. Explicitly, since we have already analytically show that the cost is $O(N_T nN)$ for ICS to sample $N_T$ error-sensitive Clifford circuits with $n$ qubits and $N$ gates in the ``Sampling algorithms'' section, we here show that $N_T$ required for optimising the error mitigation formula is finite and does not increase with either the number of qubits or the number of gates. Suppose $y_{\bfC}^\prime (\lambda)$ is the error mitigation formula parameterised by $\lambda$, the optimal error mitigation formula is $y_{\bfC}^\prime (\lambda_u)$, where $\lambda_u$ minimises MSE over unitary circuits, i.e. $\lambda_u = \text{argmin}_\lambda L_\mathbb{U}^\prime (\lambda)$, and $L_\mathbb{U}^\prime(\lambda) = \langle (y_{\bfC}^\prime(\lambda)-f_{\bfC})^2\rangle_{\bfC \in \mathbb{U}}$. We aim to find $\lambda_u$ via ICS. Suppose the optimised parameter we find via ICS is $\lambda_c$. If the number of unitary circuits and Clifford circuits are both infinite, it holds that $\lambda_c = \lambda_u$ and $L_{\mathbb{U}}^\prime(\lambda_c) = L_{\mathbb{U}}^\prime(\lambda_u)$. If the number of circuits $N_T$ is finite, we should expect that $\lambda_c \neq \lambda_u$ and $L_{\mathbb{U}}^\prime(\lambda_c) \leq L_{\mathbb{U}}^\prime(\lambda_u)$ (since $\lambda_u$ always minimises $L_\mathbb{U}^\prime$). We demonstrate in Fig.~\ref{fig:costScaling} that, with a finite $N_T$, $L_\mathbb{U}^\prime(\lambda_c) / L_\mathbb{U}^\prime(\lambda_u)$ is always finite and close to 1 for both algorithms proposed in the ``Sampling algorithms'' section, which verifies that it is feasible to implement optimised error mitigation utilising ICS.

\section{Assumptions about the circuits}
\label{sec:conditions}
Having proposed algorithms for generating error-sensitive circuits with different distributions, we can use these circuits to determine the variational parameters in an error mitigation formula. The equivalence between Clifford sampling and unitary sampling is under the condition that errors are single-qubit-gate independent. When errors are weakly single-qubit-gate dependent, we can use hybrid sampling to estimate $L_{\bbU}$~\cite{wang_scalable_2021}. Clifford-dominant circuits are circuits with a few non-Clifford gates, and they can also be efficiently simulated on a classical computer. Hybrid sampling uses both Clifford and Clifford-dominant circuits. The equivalence between ICS and unitary sampling holds for Pauli error models. When there are errors other than Pauli errors occurring in the circuit, non-sensitive Clifford circuits with $f_{\bfC} = 0$ may also respond to errors. In this case, we can sample some non-sensitive circuits, maybe with a smaller probability than error-sensitive circuits according to the principle of importance sampling, to estimate $L_{\bbU}$.

Our methods rely on the two-qubit gates in the circuit being Clifford gates and can be generalised to situations when two-qubit gates are non-Clifford in practice. We take the Molmer-Sorensen (MS) gate~\cite{sorensen_quantum_1999} $MS(\theta)=\exp(i\theta Z_1 Z_2)$ as an example to illustrate the generalisations. The methods can be generalized as long as we can transpile the circuit to a two-qubit Clifford circuit. For example, instead of directly implementing the MS gate at the physical level, one can implement it via $MS(\theta)=(\mathrm{CNOT})\exp(i\theta Z_2)(\mathrm{CNOT})$ where CNOT gate is realised with MS gate and single-qubit gates. Although the transpiled gate sequence introduces extra errors since we used two MS gates to realize one, it may be worthwhile considering the significant error reduction using our method. Without explicitly implementing the transpilation above, there are other approaches to apply the method depending on the practical condition. If the error is independent of $\theta$, we can randomly choose $\theta$ from $\{0, \pi/2, \pi, 3\pi/2\}$ such that the training circuit is Clifford. If the error weakly depends on $\theta$, we can employ the hybrid sampling mentioned above, i.e. we allow a few $\theta$ to be arbitrary and sample the others from $\{0, \pi/2, \pi, 3\pi/2\}$. In some restricted situations, the method can be applicable regardless of the dependence of error on $\theta$. For example, in trotterised circuits or unitary coupled cluster circuits, there are many repeated blocks of the form $B(\theta) = (\prod_{i=1}^n S_i)\exp(i\theta\prod_{i=1}^n P_i)$, where $S_i$ is a single-qubit unitary gate and $P_i$ is a Pauli operator on the $i$-th qubit. If the noise only depends on the length of pulse manipulating qubits, i.e. the absolute value of $\theta$, one can let $\theta = -\theta_\ast$ and $S_i=I$ for the block right before a block where $\theta=\theta_\ast$, which makes the circuit Clifford.

\begin{figure}
    \begin{center}
        \includegraphics[scale=0.5]{\figpath/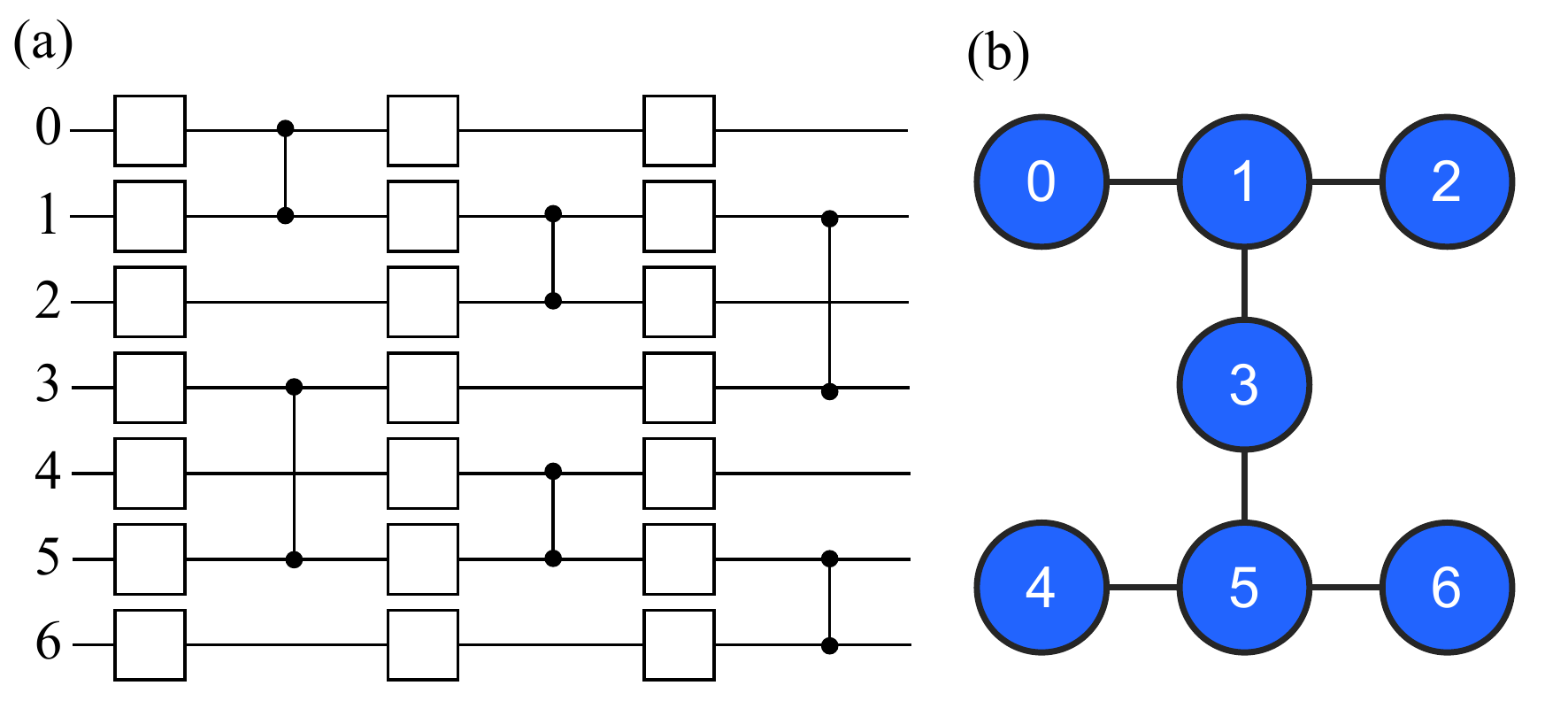}
        \caption{Circuit frame and qubit network regarding the 7-qubit experiments. (a) The repeating block of the circuit frame used in the experiments on 7-qubit \textit{ibm\_nairobi}. (b) Qubit network of \textit{ibm\_nairobi}.}
        \label{fig:frame_nairobi}
    \end{center}
\end{figure}

\begin{figure*}
    \begin{center}
        \includegraphics{\figpath/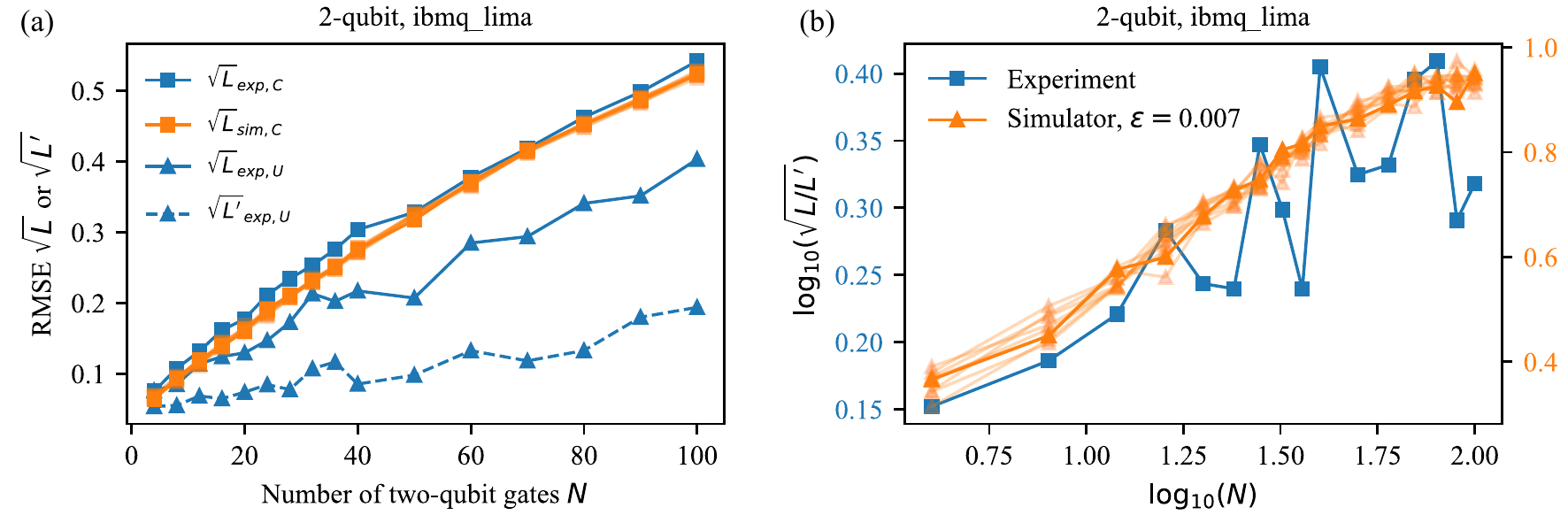}
        \includegraphics{\figpath/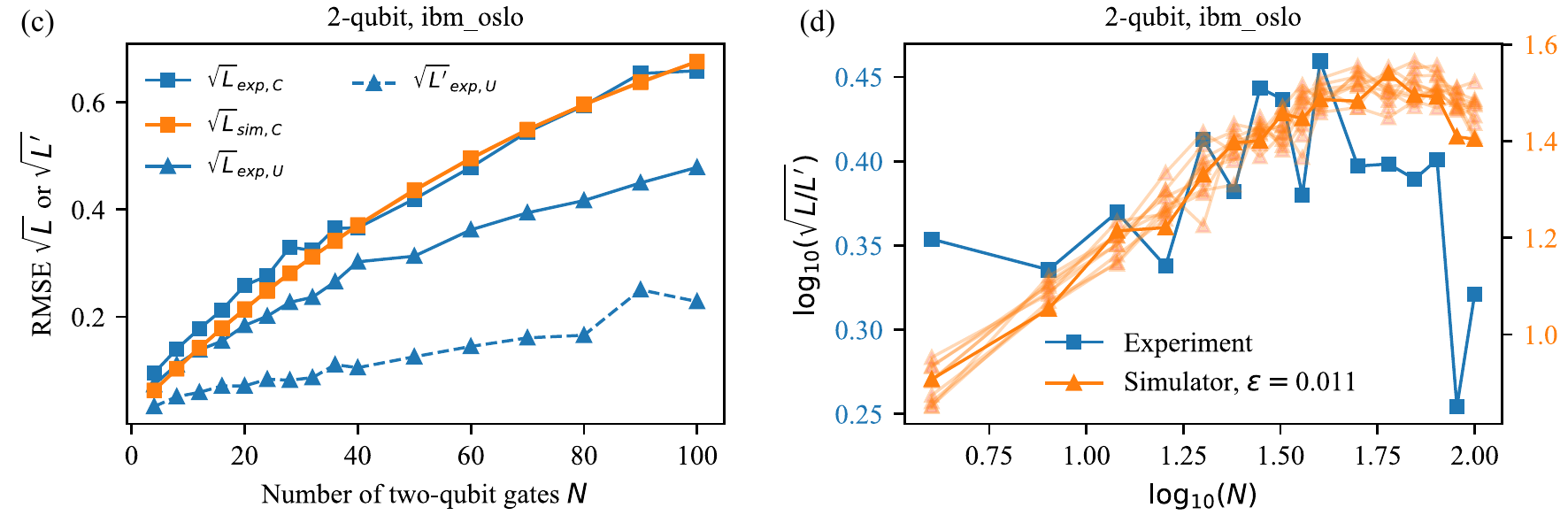}
        \caption{Two-qubit experimental results on \textit{ibmq\_lima} and \textit{ibm\_oslo}. In the numerical simulations for comparison, the error rates per CNOT gate are $\epsilon = 0.007$ and $\epsilon = 0.011$, respectively. In (a) and (c), $\sqrt{L}_{exp,C}$ and $\sqrt{L}_{sim,C}$ are RMSEs of Clifford circuits without error mitigation obtained in experiments and simulations, respectively, and $\sqrt{L}_{exp,U}$ and $\sqrt{L^\prime}_{exp,U}$ are RMSEs of general unitary circuits without and with error mitigation, respectively, obtained in experiments. In (b) and (d), the error suppression ratios $\sqrt{L/L^\prime}$ are plotted, which are computed using general unitary circuits. Axes on the left and right sides correspond to experimental and numerical results, respectively. The translucent curves show numerical simulations with different randomly generated circuits.}
        \label{fig:expResult2}
    \end{center}
\end{figure*}

\begin{figure*}
    \begin{center}
        \includegraphics{\figpath/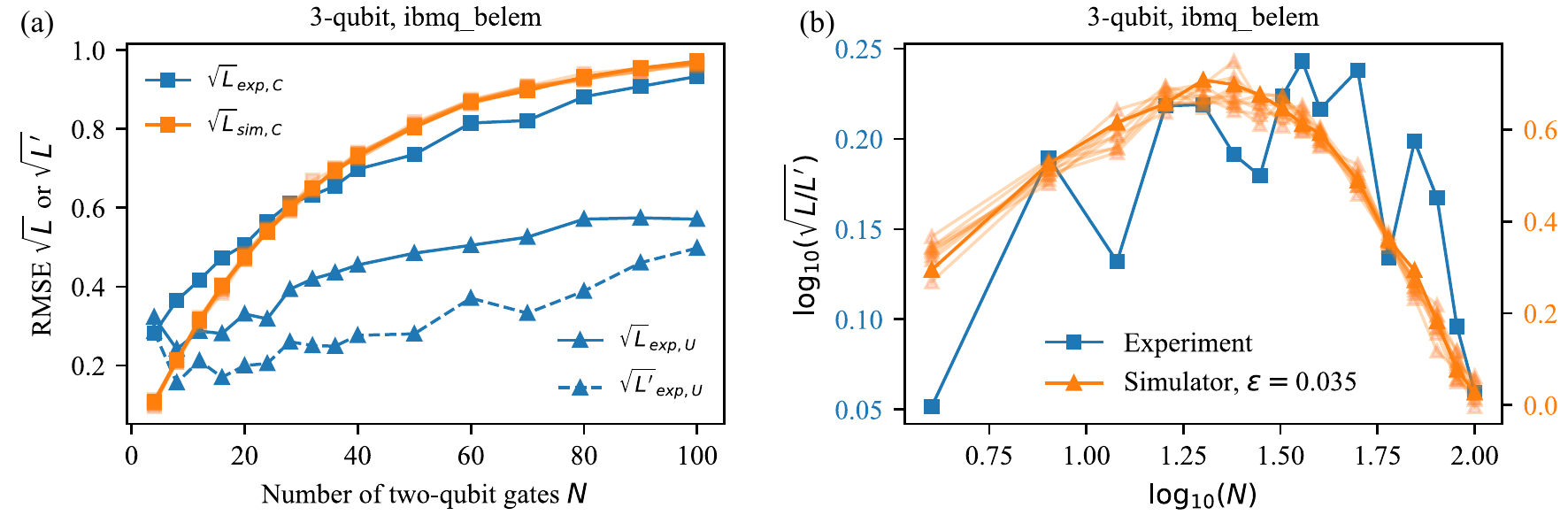}
        \includegraphics{\figpath/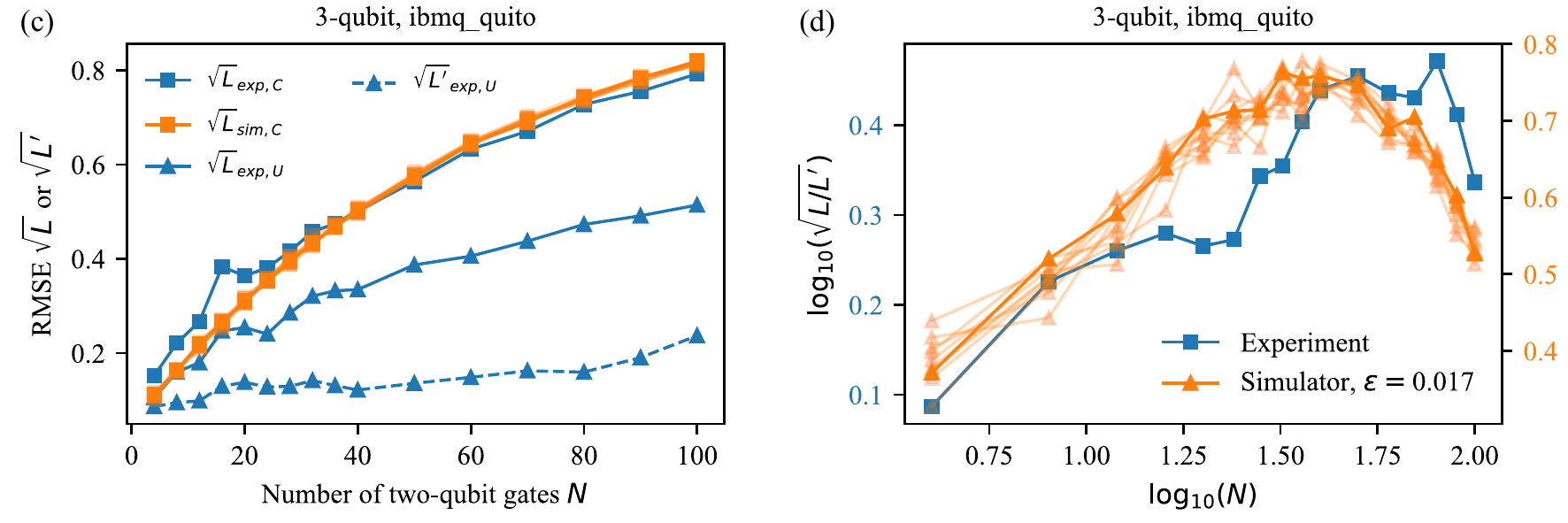}
        \caption{Three-qubit experimental results on \textit{ibmq\_belem} and \textit{ibmq\_quito}. In the numerical simulations for comparison, the error rates per CNOT gate are $\epsilon = 0.035$ and $\epsilon = 0.017$, respectively. In (a) and (c), $\sqrt{L}_{exp,C}$ and $\sqrt{L}_{sim,C}$ are RMSEs of Clifford circuits without error mitigation obtained in experiments and simulations, respectively, and $\sqrt{L}_{exp,U}$ and $\sqrt{L^\prime}_{exp,U}$ are RMSEs of general unitary circuits without and with error mitigation, respectively, obtained in experiments. In (b) and (d), the error suppression ratios $\sqrt{L/L^\prime}$ are plotted, which are computed using general unitary circuits. Axes on the left and right sides correspond to experimental and numerical results, respectively. The translucent curves show numerical simulations with different randomly generated circuits.}
        \label{fig:expResult3}
    \end{center}
\end{figure*}

\begin{figure*}
    \begin{center}
        \includegraphics{\figpath/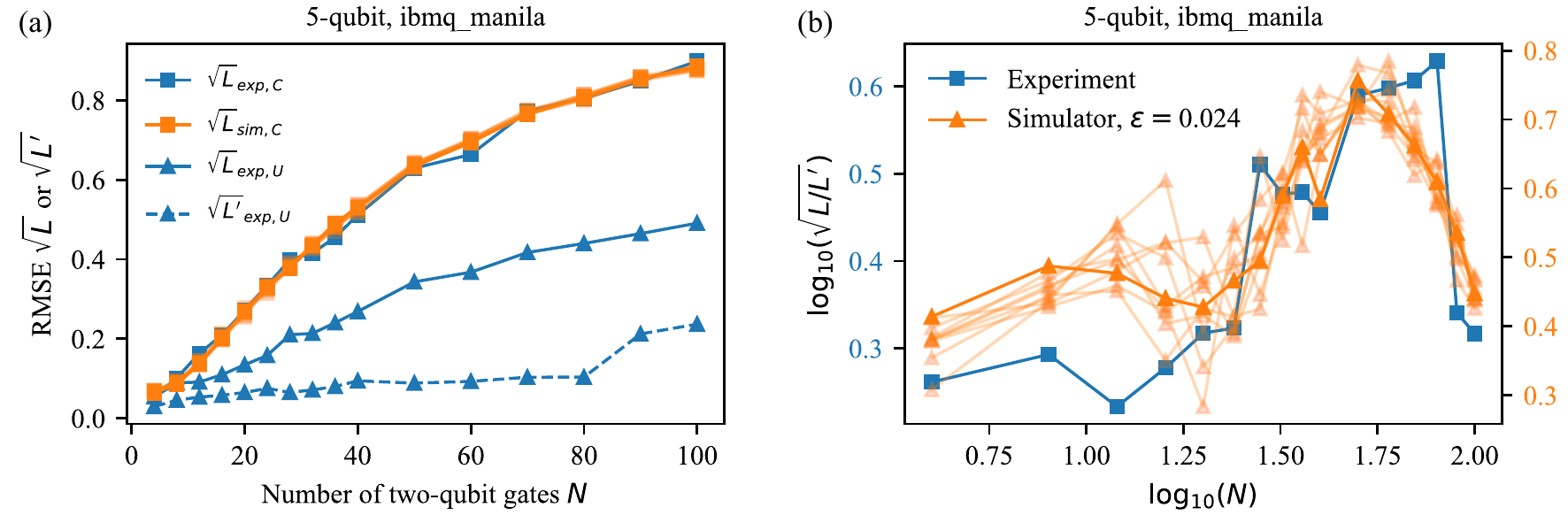}
        \includegraphics{\figpath/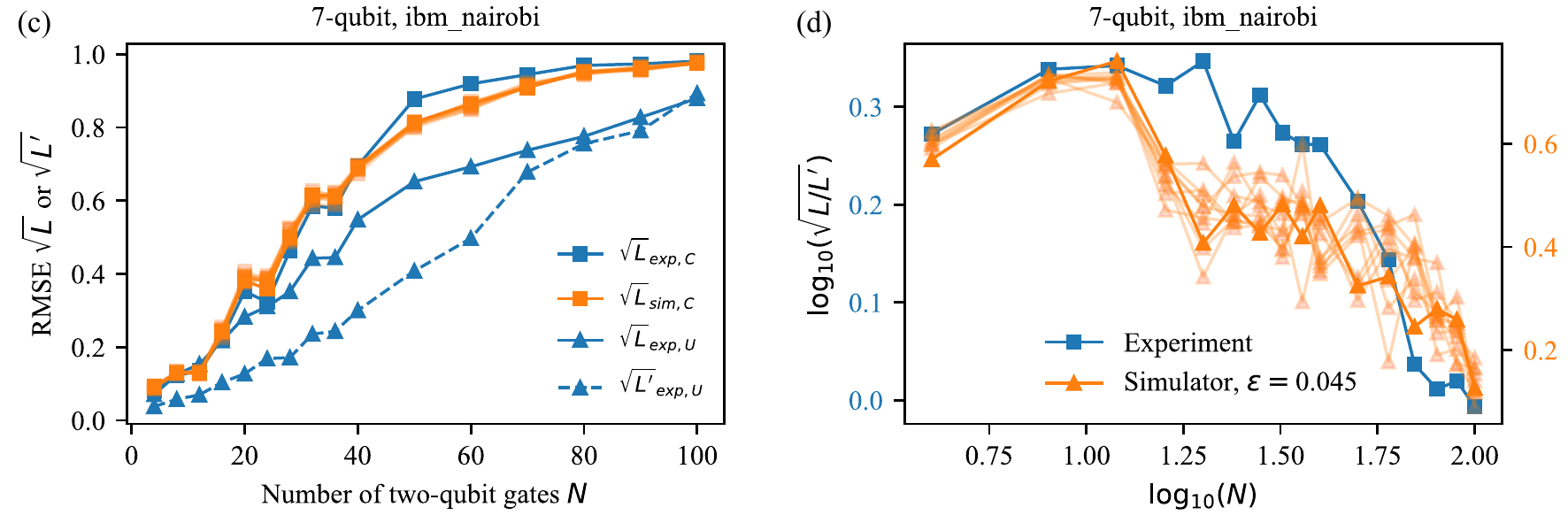}
        \caption{Five-qubit and seven-qubit experimental results on \textit{ibmq\_manila} and \textit{ibm\_nairobi}. In the numerical simulations for comparison, the error rates per CNOT gate are $\epsilon = 0.024$ and $\epsilon = 0.040$, respectively. In (a) and (c), $\sqrt{L}_{exp,C}$ and $\sqrt{L}_{sim,C}$ are RMSEs of Clifford circuits without error mitigation obtained in experiments and simulations, respectively, and $\sqrt{L}_{exp,U}$ and $\sqrt{L^\prime}_{exp,U}$ are RMSEs of general unitary circuits without and with error mitigation, respectively, obtained in experiments. In (b) and (d), the error suppression ratios $\sqrt{L/L^\prime}$ are plotted, which are computed using general unitary circuits. Axes on the left and right sides correspond to experimental and numerical results, respectively. The translucent curves show numerical simulations with different randomly generated circuits. In the seven-qubit experiment, we use the same method as in Appendix~\ref{sec:appDepFc} to generate random unitary circuits. }
        \label{fig:expResult57}
    \end{center}
\end{figure*}

\section{Dependence on the error-free computation result}
\label{sec:appDepFc}

In the ``Error scaling in optimised error mitigation formulas'' section, we have demonstrated that the optimised error mitigation can significantly reduce RMSE. Such a result is obtained with completely random unitary circuits, in which the typical outcome is close to zero. It is natural to ask whether the error suppression ratio obtained via averaging random unitary circuits still applies to atypical but useful circuits such as the circuits in variational quantum eigensolver and quantum approximation optimisation algorithms. Here, we numerically investigate the dependence of the error suppression ratio on the error-free expected value. The result is shown in Fig.~\ref{fig:fc_dependence}. We can find that the error suppression ratio for the atypical unitary circuits (that the error-free outcome is far from zero) is close to that obtained via averaging random unitary circuits. From the figure, we can also see that the error suppression ratio is relatively small when the error-free outcome is close to zero; This is not problematic since the absolute error is usually small when the error-free outcome is small.

Because of the low probability of successfully finding a circuit with near-one error-free outcome via post-selecting completely random unitary circuits, we take a trick to overcome this problem. We first randomly generate error-sensitive Clifford circuits using ICS, and then we alter single-qubit gates with small-angle random rotations, such that random unitary circuits with near-one outcomes can be efficiently generated.

\section{Revisiting the scalability and validation on quantum processors}
\label{app:experiment}

Our theoretical analysis and numerical simulations are mainly based on the gate depolarising model, which predicts that the error suppression ratio $\sqrt{L/L^\prime}$ increases with the gate number as $\propto\sqrt{N}$. Here, we experimentally verify this on six IBM quantum computers. In all experiments, we observe that the error suppresion ratio $\sqrt{L/L^\prime}$ increases in the regime of low total error rate, and decreases when the gate number is too large, which is consistent with the numerical result of the gate depolarising model. We note that the decrease is expected, since it occurs when the condition of modest total error rate is vioalted (see the ``Phenomenological error model'' section for the discussion of this condition) and can be explained by the gate depolarising error model.

In the experiments, we take the PEMI protocol in error mitigation. The experiments are performed on six IBM Quantum open-access devices~\cite{ibmq} with various numbers of qubits and numbers of CNOT gates. We perform two-qubit experiments on \textit{ibmq\_lima} and \textit{ibm\_oslo}, three-qubit experiments on \textit{ibmq\_belem} and \textit{ibmq\_quito}, a five-qubit experiment on \textit{ibmq\_manila} and a seven-qubit experiment on \textit{ibm\_nairobi}. The number of CNOT gates ranges from four to a hundred. For each pair of qubit number and gate number, we randomly generate 50 error-sensitive Clifford circuits via Algorithm~\ref{alg:nonU} to determine the optimal parameter in the PEMI protocol and generate 50 unitary circuits with error-free outcomes larger than 0.5 to compute the RMSEs $\sqrt{L}$ and $\sqrt{L^\prime}$. Each circuit is measured for 100000 shots. For two-, three- and five-qubit experiments, the circuit frame is the same as that in Fig.~\ref{fig:frames}(a), without the two-qubit gate acting on the first and last qubits because there is no direct connection between these two qubits on the devices. For the seven-qubit experiment on \textit{ibm\_nairobi}, the circuit frame is depicted in Fig.~\ref{fig:frame_nairobi}(a) due to the qubit network shown in Fig.~\ref{fig:frame_nairobi}(b). The qubit network only allows up to two CNOT gates implemented in parallel (which may cause significant idle-operation errors compared with other experiments). The experimental results are shown in Figs.~\ref{fig:expResult2}, \ref{fig:expResult3} and \ref{fig:expResult57}.

As a comparison to the experiments, we simulate the same circuits numerically with the gate depolarising error model. In addition to gate errors, measurement errors are also taken into account. We take the measurement error rate provided by IBM Quantum servers. In gate operations, we assume that only CNOT gates have errors, i.e.~the error rate per CNOT gate effectively includes errors in single-qubit gates and idle operations. Notice that the gate depolarising model is different from the actual error model of the devices. Therefore, instead of taking the CNOT-gate error rate provided by IBM Quantum servers, we determine the error rate by fitting the numerical result to the experimental result. Notice that in fitting, we only use the experimental result of Clifford circuits. Specifically, blue squares in the left panels in Figs.~\ref{fig:expResult2}, \ref{fig:expResult3} and \ref{fig:expResult57} represent experimental data of Clifford circuits, denoted by $\sqrt{L}_{exp,C}(N)$, where $N$ is the gate number; orange squares represent numerical data, denoted by $\sqrt{L}_{sim,C}(N,\epsilon)$, where $\epsilon$ is the error rate per CNOT gate; and we determine the value of $\epsilon$ by minimising $\sum_{N}[\sqrt{L}_{exp,C}(N) - \sqrt{L}_{sim,C}(N,\epsilon)]^2$. The values of $\epsilon$ can be found in figure captions for each experiment. With the error rates, we repeat each experiment on the numerical simulator ten times: For one of them, we take the same random circuits used in the experiment, and the result is represented by orange triangles in the right panels in Figs.~\ref{fig:expResult2}, \ref{fig:expResult3} and \ref{fig:expResult57}; and for the other nine simulations, we regenerate random circuits, and the results are represented by translucent orange triangles. The disparity among the orange curves suggests moderate fluctuation due to the finite number of random circuits.

We can find that the experimental behaviour of the RMSE ratio $\sqrt{L/L^\prime}$ is consistent with the numerical result. The main result of this work is that $\sqrt{L/L^\prime}$ increases with the gate number as $\propto \sqrt{N}$ under the condition that the total error rate is modest. Given a constant error rate per gate, a large gate number violates the condition and causes a vanishing fidelity. In both the experimental and numerical results, $\sqrt{L/L^\prime}$ increases with $N$ then decreases when $N$ is too large. In each experiment, the turning points occur at similar gate numbers in the experiment and numerical simulation. Notice that $\sqrt{L/L^\prime}$ is the RMSE ratio for general unitary circuits rather than Clifford circuits, therefore, the consistent behaviour is not due to the fitting (The fitting is implemented for Clifford circuits to work out the error rate per gate). Actually, the absolute values of the RMSE ratio are different in the experiment and numerical simulation, though their trends are the same. This difference is reasonable because the gate depolarising model is different from the actual error models of devices.

\clearpage
\bibliographystyle{MyBst.bst}
\bibliography{icsem_merged.bbl}
\end{document}